\documentclass[journal]{IEEEtran}
\pdfoutput=1

\usepackage[cmex10]{amsmath}
\usepackage{amssymb}
\usepackage{cite}
\usepackage[mathcal]{euscript}
\usepackage{float}
\usepackage{graphicx}
\usepackage{nicefrac}
\usepackage{subfigure}

\newtheorem{theorem}{Theorem}[section]
\newtheorem{example}[theorem]{Example}
\newtheorem{lemma}[theorem]{Lemma}
\newtheorem{proposition}[theorem]{Proposition}

\newcommand\Proof{\phantom{x}$\qquad${\it Proof: }}
\newcommand\Remark{\vspace*{0.05in}\noindent{\textbf{Remark:} }}
\newcommand\Remarks{\vspace*{0.05in}\noindent{\textbf{Remarks:} }}
\newcommand\nc{\newcommand}
\nc\etal    {~\emph{et~al}.~}
\nc\floor[1]{\left\lfloor{#1}\right\rfloor}
\nc\ceil[1] {\lceil{#1}\rceil}
\nc\half    {\nicefrac12}
\nc\ip[1]   {\langle{#1}\rangle}
\nc\norm[1] {\|{#1}\|}
\nc\remove[1]{}
\nc\qed{\rule{2mm}{2mm}\medskip}

\nc\bfa{{\boldsymbol a}}\nc\bfA{{\bf A}}\nc\cA{{\mathcal A}}\nc\sA{{\mathscr A}}
\nc\bfb{{\boldsymbol b}}\nc\bfB{{\bf B}}\nc\cB{{\mathcal B}}\nc\sB{{\mathscr B}}
\nc\bfc{{\boldsymbol c}}\nc\bfC{{\bf C}}\nc\cC{{\mathcal C}}\nc\sC{{\mathscr C}}
\nc\bfd{{\boldsymbol d}}\nc\bfD{{\bf D}}\nc\cD{{\mathcal D}}\nc\sD{{\mathscr D}}
\nc\bfe{{\boldsymbol e}}\nc\bfE{{\bf E}}\nc\cE{{\mathcal E}}\nc\sE{{\mathscr E}}
\nc\bff{{\boldsymbol f}}\nc\bfF{{\bf F}}\nc\cF{{\mathcal F}}\nc\sF{{\mathscr F}}
\nc\bfg{{\boldsymbol g}}\nc\bfG{{\bf G}}\nc\cG{{\mathcal G}}\nc\sG{{\mathscr G}}
\nc\bfh{{\boldsymbol h}}\nc\bfH{{\bf H}}\nc\cH{{\mathcal H}}\nc\sH{{\mathscr H}}
\nc\bfi{{\boldsymbol i}}\nc\bfI{{\bf I}}\nc\cI{{\mathcal I}}\nc\sI{{\mathscr I}}
\nc\bfj{{\boldsymbol j}}\nc\bfJ{{\bf J}}\nc\cJ{{\mathcal J}}\nc\sJ{{\mathscr J}}
\nc\bfk{{\boldsymbol k}}\nc\bfK{{\bf K}}\nc\cK{{\mathcal K}}\nc\sK{{\mathscr K}}
\nc\bfl{{\boldsymbol l}}\nc\bfL{{\bf L}}\nc\cL{{\mathcal L}}\nc\sL{{\mathscr L}}
\nc\bfm{{\boldsymbol m}}\nc\bfM{{\bf M}}\nc\cM{{\mathcal M}}\nc\sM{{\mathscr M}}
\nc\bfn{{\boldsymbol n}}\nc\bfN{{\bf N}}\nc\cN{{\mathcal N}}\nc\sN{{\mathscr N}}
\nc\bfo{{\boldsymbol o}}\nc\bfO{{\bf O}}\nc\cO{{\mathcal O}}\nc\sO{{\mathscr O}}
\nc\bfp{{\boldsymbol p}}\nc\bfP{{\bf P}}\nc\cP{{\mathcal P}}\nc\sP{{\mathscr P}}
\nc\bfq{{\boldsymbol q}}\nc\bfQ{{\bf Q}}\nc\cQ{{\mathcal Q}}\nc\sQ{{\mathscr Q}}
\nc\bfr{{\boldsymbol r}}\nc\bfR{{\bf R}}\nc\cR{{\mathcal R}}\nc\sR{{\mathscr R}}
\nc\bfs{{\boldsymbol s}}\nc\bfS{{\bf S}}\nc\cS{{\mathcal S}}\nc\sS{{\mathscr S}}
\nc\bft{{\boldsymbol t}}\nc\bfT{{\bf T}}\nc\cT{{\mathcal T}}\nc\sT{{\mathscr T}}
\nc\bfu{{\boldsymbol u}}\nc\bfU{{\bf U}}\nc\cU{{\mathcal U}}\nc\sU{{\mathscr U}}
\nc\bfv{{\boldsymbol v}}\nc\bfV{{\bf V}}\nc\cV{{\mathcal V}}\nc\sV{{\mathscr V}}
\nc\bfw{{\boldsymbol w}}\nc\bfW{{\bf W}}\nc\cW{{\mathcal W}}\nc\sW{{\mathscr W}}
\nc\bfx{{\boldsymbol x}}\nc\bfX{{\bf X}}\nc\cX{{\mathcal X}}\nc\sX{{\mathscr X}}
\nc\bfy{{\boldsymbol y}}\nc\bfY{{\bf Y}}\nc\cY{{\mathcal Y}}\nc\sY{{\mathscr Y}}
\nc\bfz{{\boldsymbol z}}\nc\bfZ{{\bf Z}}\nc\cZ{{\mathcal Z}}\nc\sZ{{\mathscr Z}}
\nc\cc  {{\mathcal C}}
\nc\ff  {\mathbb{F}}
\nc\nn  {\mathbb{N}}
\nc\rr  {\mathbb{R}}
\nc\tT  {\mathbb{T}}
\nc\integers{{\mathbb Z}}
\nc\rf  {{\mathfrak f}}
\nc\rg  {{\mathfrak g}}

\begin{document}
\title{Estimates on the Size of Symbol Weight Codes}
\author{Yeow Meng Chee,~\IEEEmembership{Senior~Member,~IEEE},
    Han Mao Kiah,~\IEEEmembership{Student~Member,~IEEE},\\
Punarbasu Purkayastha,~\IEEEmembership{Member,~IEEE}
\thanks{Y.~M.~Chee, H.~M.~Kiah, and P.~Purkayastha are with the 
    Division of Mathematical Sciences, School of Physical \&
    Mathematical Sciences, Nanyang Technological
    University, Singapore -- 637371. e-mails: \{ymchee, kiah0001,
    punarbasu\}@ntu.edu.sg.}%
\thanks{Research of the authors was supported in part by the National Research
Foundation of Singapore under Research Grant NRF-CRP2-2007-03.}%
%\thanks{Manuscript received April~11, 2012; accepted July~30, 2012.}%
%
}%  <-this % stops a space
%\markboth{Journal of \LaTeX\ Class Files,~Vol.~6, No.~1, January~2007}%
%{Shell \MakeLowercase{\textit{et al.}}: Bare Demo of IEEEtran.cls for Journals}

\maketitle
\begin{abstract}
%\boldmath
The study of codes for powerline communications has garnered much
interest over the past decade. Various types of codes such as permutation
codes, frequency permutation arrays, and constant composition codes have
been proposed over the years. In this work we study a type of code called
the bounded symbol weight codes which was first introduced by Versfeld\etal~in 2005, and a related family of codes that we term constant symbol
weight codes. We provide new upper and lower bounds on the size of bounded
symbol weight and constant symbol weight codes. We also give 
direct and recursive constructions of codes for certain parameters.
\end{abstract}

\section{Introduction}
The notion of transmitting data over powerlines has posed an interesting
challenge for information and coding theory. The noise characteristics of
such a communication channel include permanent narrowband noise, impulse
noise and white Gaussian noise. 
Communication over this channel also has an additional requirement
that the power envelope be as close to constant as possible. 
Vinck \cite{vin00} studied this channel and showed that $M$-ary
Frequency Shift Keying ($M$-FSK) modulation, in conjunction with the use of
permutation codes, provides a constant power envelope, frequency
spreading and redundancy to correct errors resulting from the
harsh noise pattern. This has since resulted in research on Frequency
Permutation Arrays (FPAs) and constant composition codes (CCCs) which retain the
property of a constant power envelope (see
\cite{che08,che10,chu04,chu06,col04, din05, din06, duk11, ges95, huc06,
huc06-2, huc10}, and \cite{huc06} for a survey).
Every codeword of an FPA or a CCC has the requirement that the frequency of
each symbol is fixed by the parameters of the code.
Versfeld\etal\cite{ver05}
introduced the notion of the ``same-symbol weight'' of a code by relaxing the
requirement that every symbol must occur a fixed number of times in any codeword. 
In every codeword of a same-symbol weight code, the frequency of any symbol
is bounded.  Even with this relaxation it is possible to detect permanent
narrowband noise. Versfeld\etal\cite{ver05,ver10} used Reed-Solomon
codes to design codes with specified same-symbol weight.

In this work we mostly study the asymptotic behavior of codes in the symbol weight space. We use the term
{\em bounded symbol weight} (as opposed to same-symbol weight 
\cite{ver05}) to denote all words in the Hamming space with bounded symbol
weight, that is, any symbol in a codeword does not occur more than a fixed
number of times, say $r.$ 
This terminology is adopted in order to distinguish this space from the
{\em constant symbol weight space}, in which every symbol in a word in the
Hamming space occurs at most $r$ times and there exists one symbol which
occurs exactly $r$ times. The constant symbol weight space is clearly
a subset of the bounded symbol weight space. 
We also use the term symbol
weight space to refer to either the bounded symbol weight or constant
symbol weight space. The actual space being referred to is made clear from
the context and notation.

As described in
\cite{ver05,ver10}, the symbol weight determines whether the code can
detect and correct narrowband noise in the powerline channel. 
An FPA or a CCC belongs to some constant symbol weight space.
The constant symbol weight space also contains other compositions all of
which have the same maximal part, that is, all such codes have the same
fixed symbol weight. Thus a code in the constant symbol weight space is
larger than a CCC of a fixed composition, and is still relevant for
correcting narrowband noise. The asymptotic behavior of FPAs have been
studied in \cite{bla79, chu06, huc06}. In contrast, there are
relatively fewer results on the asymptotic behavior of CCCs
(see \cite{sido75, luo11}). We consider familiar techniques used to
derive classical bounds such as the GV bound, the Johnson bound and the
Singleton bound, on codes in the symbol weight space.
However, the derivation of these results are not immediate
because of the lack of any reasonable structure in the symbol weight spaces.
In particular, even the Hamming balls of a fixed radius in these spaces
depend on the center of the ball. In Section \ref{sec:lower-bnd-sw}, we
also study non-asymptotic bounds on codes in the symbol weight spaces by
expressing them in terms of different CCCs. This also raises related
combinatorial questions regarding the size and construction of optimal codes,
which can be an interesting avenue of future research.
In later sections we show that there exists
codes, which are subsets of Reed-Solomon codes, with high rate and relative
distance, which are a subset of the constant symbol weight space.

Throughout this work we are mostly concerned with codes that have positive rate
and positive relative distance.  Hence we do not study codes with very
large distances, in the  Plotkin region. 
We start with some basic definitions and notations in the following
section. We devote Section \ref{sec:asymp-sws} to deriving the exact and
asymptotic size of the symbol weight spaces. The results in Section
\ref{sec:asymp-sws} allow us to determine which constant composition space
contained within the symbol weight space contributes the most to the rate
of a  symbol weight space. These estimates are used in Section
\ref{sec:sw-asymp} to determine upper and lower bounds on bounded symbol
weight and constant symbol weight codes. In particular, it is clear that
asymptotically some constant composition code determines the rate of
a symbol weight code. An upper bound is readily obtained from either the
Singleton bound or the Linear Programming bound in the Hamming space.
In Section \ref{sec:sw-asymp} we also provide a Johnson-type bound on codes
in the constant symbol weight space, and use this bound to derive an
asymptotic improvement of the Singleton bound and the Linear Programming
bound for certain ranges of the minimum distances and the symbol weight. In
Section \ref{sec:lower-bnd-sw} we provide non-asymptotic lower bounds on
symbol weight codes. We introduce a new metric on the space of compositions
of an integer and use this metric to lower bound the size of symbol weight
codes by a sum of sizes of CCCs. Finally, in Section
\ref{sec:constructions} we provide other constructions of constant symbol
weight codes, and in particular show that the asymptotic lower bound
presented in Section \ref{sec:asymp-sws} is tight for certain
parameters, for subcodes of Reed-Solomon codes.

\section{Preliminaries}
Let $\integers_q = \{0,\dots,q-1\}$ denote an alphabet set of $q$ elements.
We consider symbol weight codes in the Hamming space $\integers_q^n
= \{0,\dots,q-1\}^n.$
The \emph{symbol weight} of a word is defined as the maximum of the 
frequencies of occurrences of symbols in the word.
For instance, the all-0 word has a symbol weight of $n.$
The \emph{bounded symbol weight space} with symbol weight $r$ is the set of
all words with symbol weight at most $r.$ This space is denoted by
$SW(n,q,\le{r})$.
The bounded symbol weight space is termed as ``same-symbol weight space''
in the works of Versfeld\etal\cite{ver05,ver10}. We adopt 
this terminology to distinguish this space from the constant symbol weight
space that we define next.
In the \emph{constant symbol weight space}
every word has a symbol weight of exactly $r.$ This space is
denoted as $SW(n,q,r)$. 
If every symbol occurs in each codeword we can use the Pigeonhole principle
to get the lower bound 
$r\ge \ceil{n/q}$. Since any word with this lowest value
of symbol weight contains the least repetition of any symbol, these words
are considered as ones with the \emph{optimal} symbol weight.

In this work we study codes in the  bounded and constant symbol weight
spaces.  A bounded (resp. constant) symbol weight code is a subset of the
bounded (resp. constant) symbol weight space.
Let $A_q^{SW}(n,d,\le{r})$ (resp. $A_q^{SW}(n,d,{r})$) denote the
maximum size of a bounded (resp. constant) symbol weight code with distance
$d$ in $SW(n,q,\le{r})$ (resp. $SW(n,q,{r})$). We denote a composition of
$n$ into $q$ non-negative parts 
by $\bfn = [n_0,n_1,\dots,n_{q-1}]$. The constant composition space
with composition $\bfn = [n_0,n_1,\dots,n_{q-1}]$ is a subset of
$\integers_q^n$ in every word of which the $i$-th symbol occurs exactly
$n_i$ times. A constant composition code is a subset of a constant
composition space.
We use the notation $A_q(\bfn,d)$ to  denote the maximum size of a code in
the constant composition space given by the composition $\bfn$ and minimum
distance at least $d$. We use the notation $A_q(n,d)$ to denote the maximum
size of a code with minimum distance at least $d$ in the Hamming space.
A code $\cC$ of length $n,$ size $M,$ distance $d,$ over $\integers_q$ is
denoted by $\cC(n,M,d)_q$. If $\cC$ has a constant symbol weight $r$ it is
denoted by $\cC(n,M,d,r)_q$. If $\cC$ is a linear code of dimension $k$
over a field $\ff_q$ it is denoted as $\cC[n,k,d]_q$. 

An FPA consists of vectors in which every symbol occurs a fixed number, say
$\lambda$, of times. Hence, an FPA is a CCC with
composition $\bfn=[\lambda,\dots,\lambda]$. Thus the FPA is a subset of the
constant symbol weight space with symbol weight $\lambda.$ Similarly, it
can be seen that a CCC with composition $\bfn
= [n_0,\dots,n_{q-1}]$ is a subset of the constant symbol weight space with
symbol weight $r = \max\{n_i: i=0,\dots,q-1\}$.

The coded modulation scheme introduced by Vinck \cite{vin00} for the powerline
channel considered $M$-FSK modulation along with the use of permutation
codes. The demodulator considered is a hard-decision demodulator consisting
of an envelope detector with a threshold. At every time instance, the
demodulator provides a multivalued output consisting of all the symbols
that correspond to frequencies at which the output of the envelope
detector exceeds the threshold. A narrowband noise in this context
results in the same symbol appearing at all time instances. As explained in
Versfeld\etal\cite{ver05}, a linear
code is less effective in this channel. For instance, the all-zero codeword
can not be distinguished from a narrowband noise. Hence, permutation codes,
FPAs and CCCs are more suitable for communication in this channel.
To understand why we study
the constant symbol weight space, consider the following example.\\
\begin{example}
    Consider a CCC in $\integers_4^8$ with composition $[1,1,3,3]$ and
minimum distance $d = 4$, that is suitable for correcting narrowband noise
in a powerline channel. Let $(0,1,2,2,2,3,3,3)$ be a codeword in this CCC.
Then the vector $(0,0,0,1,1,1,2,3)$ is also a suitable vector for
correcting
narrowband noise. However this vector belongs to a different constant
composition space that contains vectors with composition $[3,3,1,1]$. Both
these constant composition spaces are a subset of the constant symbol
weight space with symbol weight 3. In Section~\ref{sec:lower-bnd-sw}, we
prove that since the compositions $[1,1,3,3]$ and $[3,3,1,1]$ have distance
$4$ in a specific metric that we define later, any vector from the
constant composition space with composition $[1,1,3,3]$ will be at
a distance at least 4 from any vector of the other space with composition
$[3,3,1,1]$. Hence, we can increase the size of the code by including all
the codewords from a CCC in the latter space.\\
\end{example}

It is clear that any constant symbol weight space with symbol weight $r$
can be written as the union of different constant composition spaces, each
of which contains vectors with the same symbol weight $r$. This relation to the
constant composition space is used throughout this work.

In the next section, we first
determine the size of the constant symbol weight space and the bounded
symbol weight space. This size is then used to determine a GV-type bound on
the symbol weight spaces. Unfortunately, the expression for the size of the
symbol weight spaces is unwieldy and gives little insight into the behavior
of the lower bounds. The symbol weight spaces are also not
ball-homogeneous, that is, the size of a Hamming ball in the space depends
on the center of the ball. For example, the bounded symbol weight
space in $\integers_3^3$ with symbol weight $r$ at most 2 has 24 vectors. The
ball of radius 1 around the vector $(1,0,0)$ contains 6 vectors, namely,
$(1,0,0),$ $ (2,0,0),$ $(1,1,0)$, $(1,2,0),$ $ (1,0,1),$ $(1,0,2)$. In contrast, the ball
of radius 1 around $(2,1,0)$ contains 7 vectors, namely,
$(2,0,0),(0,1,0),(1,1,0)$, $ (2,1,0),(2,2,0),(2,1,1),(2,1,2)$. This fact makes
it difficult to state decent lower bounds. Similar comments apply to the
computation of the Hamming bound. Hence, in the following two
sections, we instead study the asymptotic behavior of the symbol weight
spaces and the rate of the corresponding symbol weight codes.

In the following section we determine the asymptotic size of the symbol
weight space. This enables us to determine which constant composition
space, contained within the symbol weight space, has the largest size.

\section{Asymptotic size of the symbol weight space}
\label{sec:asymp-sws}
%Let $SW(n,q,r)$ denote the constant symbol weight space in $\integers_q^n$ with symbol
%weight $r.$
To determine the asymptotic size of the symbol weight space we
first state the expression for the non-asymptotic case. Each vector in the
symbol weight space corresponds to a vector in some constant composition
space. Hence we introduce some basic definitions below to describe the size
of the symbol weight space. Let us denote the set of all compositions of
$n$ into $q$ non-negative parts by $\cN$, that is
$$
\cN \triangleq \left\{\bfn\in\integers^q: \bfn\ge0, \sum_{i=0}^{q-1} n_i
= n\right\},
$$
and define
\begin{align*}
\cN(r) &\triangleq \{\bfn\in\cN : \max\{n_0,\dots,n_{q-1}\} = r\},\\
\cN(\le{r}) &\triangleq \{\bfn\in\cN : \max\{n_0,\dots,n_{q-1}\} \le r\}.
\end{align*}
Let $P(N,K,R)$ denote the compositions 
of $N$ into $K$ parts, each
part bounded between $0$ and $R.$ An expression for the size of $P(N,K,R)$
is given by \cite[pp.~1037]{ges95},
   \begin{equation}
\label{eq:pnkr}
|P(N,K,R)| = \sum_i (-1)^i \binom Ki \binom{K+N-(R+1)i-1}{K-1}.
   \end{equation}
Define $k_0$ to be $k_0 \triangleq \max\{n-(r-1)q,1\}.$ 
The quantity $k_0$ corresponds to the smallest
number of symbols that can occur with frequency exactly $r$ in any vector
with symbol weight $r$.
The size of the set $\cN(r)$ is given by the following lemma.\\
\begin{lemma}
\label{lem:size-cnr}
$$
|\cN(r)| = \sum_{k=k_0}^{\floor{n/r}} \binom qk
|P(n-rk,q-k,r-1)|,
$$
and
$$
k_0 = \max\{n-(r-1)q, 1\} = \begin{cases}
    q - \Delta, & r = \ceil{\frac nq} = \frac{n+\Delta}q,\\
    1, &\text{otherwise,}
    \end{cases}
$$
for some $\Delta\equiv\Delta(n,q)$ such that $\ 0\le\Delta\le q-1$.
\end{lemma}
\begin{IEEEproof}
If a vector $\bfv\in\integers_q^n$ has a composition $\bfn$ such that exactly $k$
of the symbols in $\integers_q$
have composition $r$ in $\bfv$, then the rest of the
symbols must satisfy the inequality,
          $$
n-rk \le (q-k)(r-1).
          $$
This inequality, in conjunction with the requirement that at least one
symbol must have composition $r$, determines the value of $k_0.$ If
a composition $\bfn$ has
exactly $k$ symbols with the value $r$, then these $k$ symbols can be
chosen in $\binom qk$ ways. The rest of the elements of $\bfn$ must correspond to a
composition of $n-rk$ into $q-k$ parts, each part bounded between $0$
and $r-1.$

Note that $r$ must satisfy $r\ge\ceil{n/q}.$ We have
\begin{align*}
n-(r-1)q \ge 1 \quad
\Leftrightarrow \quad r \le \frac{n+q-1}q.
\end{align*}
There is exactly one integer $r$ which satisfies $\ceil{n/q}\le r\le
(n+q-1)/q$. This value of $r$ is given by $r=(n+\Delta)/q$, for some
$\Delta$ such that
$0\le\Delta\le q-1$.
%\hspace{\fill}\qed\\
\end{IEEEproof}
\vspace{1mm}
The size of the constant symbol weight space $SW(n,q,r)$ can now be determined to be
\begin{multline}
\label{eq:size-swc}
|SW(n,q,r)| = \sum_{k=k_0}^{\floor{n/r}} \binom qk \binom
n{r,...,r,n-rk} \times \\
        \sum_{\bfx\in P(n-rk,q-k,r-1)} \binom
{n-rk} {x_1,\dots,x_{q-k}},
\end{multline}
where $\bfx = (x_1,\dots,x_{q-k})$, and $r$ is repeated $k$ times in the
multinomial coefficient $\binom{n}{r,\dots,r,n-rk}$. The size of the bounded symbol
weight space is a sum of the sizes of the different constant symbol weight
spaces, as shown below:
\begin{align}
|SW(n,q,\le{r})| &= \sum_{s=\ceil{n/q}}^r \sum_{k=k_0(s)}^{\floor{n/s}} \binom qk \binom
n{s,...,s,n-sk} \times\nonumber \\
        &\ \sum_{\bfx\in P(n-sk,q-k,s-1)} \binom
{n-sk} {x_1,\dots,x_{q-k}}\nonumber\\
        &= \sum_{\bfy\in P(n,q,r)} \binom n{y_1,\dots,y_q},
\label{eq:size-swc-bounded}
\end{align}
where $\bfx = (x_1,\dots,x_{q-k})$, $\bfy=(y_1,\dots,y_q)$, and $k_0(s)
= \max\{n-(s-1)q,1\}$.

The expressions in the equations above can
be used to provide GV-type existence bounds on symbol weight codes. A GV
bound on the size of a code $\cc$ with minimum distance $d$ in a space
$\mathcal S$ is given as
$$
|\cc| \ge \frac{|\mathcal S|}{V(\mathcal S, d-1)},
$$
where $V(\mathcal S, d-1)$ is the volume of a ball of radius $d-1$ in the
space $\mathcal S$. Although the sizes of the 
constant and bounded symbol weight spaces
are given by the above equations \eqref{eq:size-swc} and
\eqref{eq:size-swc-bounded}, respectively, there are several hurdles in
applying the GV-type bound directly. First the space itself lacks any
suitable structure and is not even ball-homogeneous.
Even for the special case of an FPA in which all the symbols occur equally
often in every vector, the expression for the GV (and also the Hamming bound)
is quite unwieldy because the size of the ball does not have a nice form;
see Huczynska \cite[Theorem~2.7]{huc06-2}.
Secondly, the
expressions for the sizes of the spaces are not suitable for expressing the
bound in a simple form. We instead study the asymptotic form of this bound
in the next section.
To determine the asymptotic results, we first need to
understand the behavior of the sizes of the symbol weight spaces for large
block length $n.$

The expression for the asymptotic size of the constant symbol weight space
is given by the following theorem. A similar expression for the bounded
symbol weight space can be  readily derived from this theorem, and is
mentioned at the end of this section.
The following theorem holds for any $q$ such that $q$ 
grows at most proportional to $n.$
Note that all the asymptotics are with
respect to $n$ and so the term $o(1)$ below goes to zero as $n$ goes to
$\infty$.\\
\begin{theorem}
\label{thm:sw-asymp}
For any $q,$ such that $q=\theta n^\epsilon,$ where $\theta$ is a positive
constant and $0\le\epsilon\le1$,
$$
\frac1n \log_q|SW(n,q,r)| = \begin{cases}
    h_q\left(1-\frac rn\right) - o(1), & r > \ceil{\frac nq},\\
    1 - o(1), & r = \ceil{\frac nq}.
    \end{cases}
$$
\end{theorem}
We first give a brief outline of the proof of this theorem. As mentioned
earlier, a constant symbol weight space with symbol weight $r$ is a union
of several constant composition spaces, each of which contains vectors of
symbol weight $r$. We first show in Lemma \ref{lem:nr-size} that the number
of constant composition spaces does not contribute to the rate of the
constant symbol weight space. This is not surprising and it is true even
for the Hamming space, when considered as a union of constant composition
spaces. Because of this Lemma, we now know that there is a constant
composition space which dominates the expression for the rate. Lemma
\ref{lem:multinomial} and \ref{lem:monotone} below help us determine this
dominant term. It turns out that this dominant term comes from the constant
composition space which has exactly $k_0$ symbols that occur exactly $r$
times in any vector.

We continue with the proof of the theorem, by first establishing a sequence of lemmas presented
below.
Let $h_p(x)$ be the $p$-ary entropy function defined in the range
$0\le x\le1$, as
$$
h_p(x) \triangleq -x\log_p \frac x{p-1} -(1-x) \log_p (1-x).
$$
\begin{lemma}
\label{lem:nr-size}
$$
\frac1n \log_q |\cN(r)| = o(1).
$$
\end{lemma}
\begin{IEEEproof}
The number of terms in the summation over the range $k_0\le k\le \floor{n/r}$
is at most $n.$ Hence only one of the terms in the summation dominates in
the asymptotics.
We note that  $|P(n-rk,q-k,r-1)|\le |\cN|$.
Also, $|\cN| = \binom{n+q-1}{q-1}$ (see \cite[pp.~415]{sha67}).
For a constant $q$, it shows that $P(n-rk,q-k,r-1)$ grows at most
polynomially in $n$ and hence
\begin{align*}
\frac 1n\log_q |\cN(r)| &= \frac 1n \log_q |P(n-rk,q-k,r-1)| + o(1)\\
        &\le  \frac 1n \log_q ( a n^{q-1}) + o(1)\\
        &= o(1),
\end{align*}
for some positive constant $a$. For $q=\theta n^\epsilon,\
0<\epsilon\le1$, and positive constant $\theta$, we get
\begin{equation*}
\frac1n \log_q \binom qk \le \frac qn h_2\left( \frac kq\right)\log_q 2
    + o(1) = o(1),
\end{equation*}%
and hence
\begin{multline*}
\frac 1n \log_q |P(n-rk,q-k,r-1)| \le \frac {n+q-1}n\times \\
        h_2\left( \frac{q-1}{n+q-1}\right)\log_q 2 + o(1)
        = o(1).
\end{multline*}
\end{IEEEproof}
By the above lemma, we can conclude that in the asymptotics of
(\ref{eq:size-swc}) only one term $\binom {n-rk}{x_1,\dots,x_{q-k}}$ from
the inner summation dominates in the asymptotics, and similarly only
one term from $\binom qk \binom n{r,\dots,r,n-rk}$ is present in the
asymptotics.  The dominant multinomial terms are given by an optimal
choice of $k.$
First, we determine the dominating multinomial term from
the inner summation in (\ref{eq:size-swc}). We use the following lemma.
Let $\Gamma(x)$ denote the Gamma function,
    $$
\Gamma(x) = \int_0^\infty t^{x-1} e^{-t} dt.
    $$
In particular, for an integer $x$, $\Gamma(x) = (x-1)!$.\\
\begin{lemma}\cite[pp.~109]{mar11}
\label{lem:multinomial}
Let $x_1,\dots,x_K$ be non-negative real numbers. Then
$$
\prod_{i=1}^K \Gamma(x_i) \ge \left( \Gamma\left(
            \frac{\sum_i x_i}K\right)\right)^K.
$$
\end{lemma}
This lemma immediately implies that
$
\binom N{x_1,\dots,x_K} \le \binom N{N/K,\dots,N/K}.
$
Hence the dominating term in the inner summation in (\ref{eq:size-swc}) is given by $\binom
{n-rk}{l,\dots,l},$ where $l = (n-rk)/(q-k).$\footnote{
We ignore the fact that the ratios may
not be integers. This argument can be made more rigorous, but cumbersome, by
taking the composition to be $l_0 = \floor{\frac{n-rk}{q-k}}$ for
$(q-k)(1-\{\frac{n-rk}{q-k}\})$ times and $l_1 = \ceil{\frac{n-rk}{q-k}}$
for $(q-k)\{\frac{n-rk}{q-k}\}$ times, where $\{x\}$ denotes the fractional part
    of a real number $x$. }
For large $n$ we obtain the
following asymptotic expression for $|SW(n,q,r)|$:
\begin{align}
\frac1n \log_q |SW(n,q,r)| &= \log_q n - k \frac rn \log_q
                r - \frac{n-rk}n\times\nonumber\\
    &\quad \log_q(n-rk) + \frac{n-rk}n\log_q (q-k) + \nonumber\\
    &\quad \frac 1n\log_q \binom qk - o(1).
\label{eq:size-swc-log}
\end{align}
Neglecting the $o(1)$ term, the maximum of the expression in equation
\eqref{eq:size-swc-log} over $k$ yields the rate of the constant symbol
weight space. Unfortunately, a closed-form expression for the optimizing
value of $k$ seems difficult to achieve, even if $k$ is considered over
reals instead of integers. We instead look at how the expression behaves
for large $n.$ The lemma below asserts that the maximum is achieved at $k^*
= k_0$ as $n \to \infty.$\\

\begin{lemma}
\label{lem:monotone}
Let $\{f_n(x)\}_{n=1}^{\infty}$ be a family of bounded, strictly monotonic
decreasing functions in $x$, defined over the domain $x\in[x_0,X_0]$, such
that $\lim_{n\to\infty} f_n(x) = f(x)$. Let
$\{g_n(x)\}_{n=1}^{\infty}$ be a family of non-negative functions such that
$0 \le g_n(x)\le C_n,$ where $C_n$ depends
only on $n$ and $\lim_{n\to\infty}C_n = 0$. Then,
\begin{multline*}
\max_{x\in [x_0,X_0]} \lim_{n\to\infty} f_n(x)+g_n(x)  = f(x_0)\\
        = \lim_{n\to\infty}
\max_{x\in [x_0,X_0]} f_n(x) + g_n(x).
\end{multline*}
\end{lemma}
%\begin{IEEEproof}
\Proof
The strict monotonicity $f_n(x) > f_n(y)$ for any $x,y,$ $ x_0\le x<y\le
X_0$, implies that
$f(x_0)\ge f(x)$ for all $x\in[ x_0,X_0]$. Now,
    \begin{align*}
\max_{x\in [x_0,X_0]} f_n(x) + C_n &\ge \max_{x\in [x_0,X_0]} f_n(x)
    + g_n(x) \ge \max_{x} f_n(x) \\
\Rightarrow f_n(x_0) + C_n &\ge \max_{x\in [x_0,X_0]} f_n(x) + g_n(x) \ge
    f_n(x_0) \\
\Rightarrow &\lim_{n\to\infty} \max_{x\in [x_0,X_0]} f_n(x) + g_n(x)
= f(x_0).
    \end{align*}
We also get
$$
\max_{x\in [x_0,X_0]} \lim_{n\to\infty} f_n(x) + g_n(x) = \max_{x\in
    [x_0,X_0]} f(x) = f(x_0).\quad\qed
$$
%\end{IEEEproof}
This lemma implies that we can determine the asymptotic optimum of
$f_n(x)+g_n(x)$ by simply taking the limit of the sequence of numbers
$f_n(x_0) + g_n(x_0)$, which converges to $f(x_0)$.\\

%\noindent\textsc{Proof of Theorem \ref{thm:sw-asymp}}:
\begin{IEEEproof}[Proof of Theorem \ref{thm:sw-asymp}]
We apply Lemma \ref{lem:monotone} as follows. Let
\begin{align*}
F_n(k) &= \log_q n - k \frac rn \log_q r - \frac{n-rk}n\log_q(n-rk)
    + \\
    &\quad \frac{n-rk}n\log_q (q-k), \\
G_n(k) &= \frac1n \log_q \binom qk,
\end{align*}
be defined over integer $k\in[k_0,\floor{n/r}]$. $G_n(k)$ can be upper 
bounded by a  term independent of $k,$
$$
G_n(k) \le \frac 1n \log_q\binom q{\floor{q/2}},
$$
and $\lim_{n\to\infty} \frac 1n \log_q \binom q{\floor{q/2}} = 0.$
We now note that for every $n\ne rq$, $F_n(k)$ is strictly monotonically
decreasing. To establish this we relax $k$ to reals and consider the
derivative $F'_n(k)$. We get 
\begin{align*}
nF'_n(k) &= -r \log_q r + r \log_q \frac{n-rk}{q-k} + r
    - \frac{n-rk}{q-k}\\
    &= r\left( \log_q \frac{n-rk}{r(q-k)} - \left( \frac{n-rk}{r(q-k)}-1
                \right)\right)\\
    &\le 0,
\end{align*}
where the last line follows because of the fact that $n-rk \le r(q-k)$, and
that $\log x \le (x-1)$ for $0<x\le 1$, with equality at $x=1$. We also
note that $n-rk < r(q-k)$  if and only if $n\ne rq.$ Hence $F_n(k)$ is
strictly monotonic decreasing for $n\ne rq.$ For $n=rq$, $F_n(k)$ is
a constant independent of $k$ and $k_0 = q$, and hence the optimal value of
$F_n(k)$ is at $k = q$. Since Lemma \ref{lem:monotone} is applicable to
$F_n(k) + G_n(k)$, we concentrate only
on determining the asymptotics of $F_n(k_0)$. For $r>\ceil{n/q}$ we get $k_0
= 1$ and 
     \begin{align*}
F_n(1) &= - \frac rn \log_q \frac rn - \Big(1-\frac rn 
        \Big)\log_q \Big(1-\frac rn\Big)  \\
        &\quad + \Big(1 - \frac rn\Big)\log_q
        (q-1) - o(1)\\
    &= h_q\Big(1-\frac rn\Big) - o(1).
     \end{align*}
For $r=\ceil{n/q} = (n+\Delta)/q$, we have $k_0 = q-\Delta = n-(r-1)q$, and
    \begin{align*}
F_n(q-\Delta) &= \frac{\Delta(r-1)}n \log_q\frac{r}{r-1} 
    - \log_q\frac {n+\Delta}{nq} \\
    &= 1 - o(1).
    \end{align*}
This proves Theorem \ref{thm:sw-asymp}.%\hspace{\fill}\qed\\
\end{IEEEproof}
\vspace{1mm}
The exponent of the asymptotic size of the bounded symbol weight space
$SW(n,q,\le r)$ is always $n(1-o(1))$ since it contains
$SW(n,q,\ceil{n/q})$.

\section{Asymptotic size of symbol weight codes}
\label{sec:sw-asymp}
In this section we provide estimates on
the rate  of symbol weight codes for
all $q=\theta n^\epsilon$, for any positive constant $\theta,$ and for
$0\le\epsilon\le1$.
We considered the asymptotic behavior of the symbol weight spaces because
of the difficulty in determining reasonable expressions for fixed $n$.
Below, we determine upper and lower bounds on the rate of a symbol weight
code. First we determine a GV-type bound 
in Theorem~\ref{thm:aqw-rate-q-n-r-const} below. 
The Singleton and Linear Programming (LP) upper bounds on codes in the Hamming
space are applicable to the symbol weight codes too.
In Theorem~\ref{thm:large-r}, we
show that for constant symbol weight codes, the Singleton and LP upper
bounds can be improved substantially for a specific range of the symbol
weight.

The following lemma is immediate and it 
shows that the rate of symbol
weight codes can be given
in terms of the rate of a CCC.\\
\begin{lemma}
\label{lem:aqw-rate}
\begin{equation}
\begin{split}
\label{eq:aqw-rate}
\frac1n \log_q A_q^{SW}(n,d,r) = 
    \frac1n \max_{\bfn\in\cN(r)} \log_q A_q(\bfn,d) + o(1),\\
\frac1n \log_q A_q^{SW}(n,d,\le{r}) = 
    \frac1n \max_{\bfn\in\cN(\le{r})} \log_q A_q(\bfn,d) + o(1).\\
\end{split}
\end{equation}
\end{lemma}
%\Proof
\begin{IEEEproof}
Note that we clearly have the following upper and lower bounds on
$A_q^{SW}(n,d,r)$:
\begin{equation}
\label{eq:aqw-bnd}
\max_{\bfn\in\cN(r)} A_q(\bfn,d)\le A_q^{SW}(n,d,r) \le |\cN(r)| \max_{\bfn\in\cN(r)} A_q(\bfn,d).
\end{equation}
The lemma now follows from an application of Lemma \ref{lem:nr-size}.
The second expression in (\ref{eq:aqw-rate}) can be determined
similarly.%\hfill\qed\\
\end{IEEEproof}
\vspace{1mm}
We state the LP upper bound on codes in the Hamming
space from Aaltonen \cite{aal90}.\\
\begin{theorem}\cite{aal90}
\label{thm:aaltonen-lp}
$$
\limsup_{n\to\infty}\frac1n\log_q A_q(n,d) \le h_q(k_q(\delta)),\quad
    0\le\delta\le\frac{q-1}q,
$$
where
$
k_q(x) = \frac{q-1}q - \frac{q-2}qx - \frac2q\sqrt{(q-1)x(1-x)},\,0\le
x\le1.
$
\end{theorem}
\vspace{2mm}

An upper bound on symbol weight codes is readily obtained
by an upper bound on codes in the Hamming space, since
   $$
A_q^{SW}(n,d,r) \le A_q^{SW}(n,d,\le{r}) \le A_q(n,d).
   $$
Thus for constant $q$ the LP bound is also
an upper bound on symbol weight codes. For $q$ growing with $n$, the
Singleton bound is an upper bound on symbol weight codes. 
Below, 
we provide asymptotic estimates of symbol weight codes.\\
\begin{theorem}
\label{thm:aqw-rate-q-n-r-const}
Let $q= \theta n^\epsilon$, where $0<\theta$ is a constant, and $0\le\epsilon\le1$.
Let $d/n \to \delta$ and  $r/n \to \rho$ as $n\to\infty$, where
$0<\delta\le \frac{q-1}q$. Then for $q$ constant, i.e., $\epsilon=0$,
\begin{align}
\liminf_{n\to\infty} \frac 1n\log_q  A_q^{SW}(n,d,r)
    &\ge h_q(1-\rho) - h_q(\delta),\nonumber \\
\liminf_{n\to\infty} \frac 1n\log_q  A_q^{SW}(n,d,\le{r})&\ge 1 - h_q(\delta).
\label{eq:lower-bound-gv}
\end{align}
For $q$ increasing with $n$ one can use the Singleton bound.
Thus for $0<\epsilon\le1$, we get
\begin{align}
    \liminf_{n\to\infty} \frac 1n\log_q  A_q^{SW}(n,d,r) &\ge
1-\rho-\delta,\quad r=\rho n,\nonumber\\
\lim_{n\to\infty} \frac 1n \log_q A_q^{SW}(n,d,r) &= 1 - \delta,\quad
r=o(n),\nonumber\\
\lim_{n\to\infty} \frac 1n \log_q A_q^{SW}(n,d,\le{r}) &= 1 - \delta,\quad
    \text{any $r$}.
\label{eq:aqw-rate-bd-r}
\end{align}
\end{theorem}
\Remark
Note that for $q$ increasing with $n$ the following limits can be inferred,
     \begin{align*}
 \lim_{n\to\infty} \frac 1n \log_q A_q^{SW}(n,d,r) = \lim_{n\to\infty}
    \frac 1n\log_q A_q(n,d), \ r = o(n),\\
 \lim_{n\to\infty} \frac 1n \log_q A_q^{SW}(n,d,\le{r}) = \lim_{n\to\infty}
   \frac 1n \log_q A_q(n,d), \ \text{any }r.
     \end{align*}
%\Proof
\begin{IEEEproof}
% NEW PROOF using Elias
We use the following lower bound on the constant symbol weight space, which is
actually an Elias-type bound on the Hamming space (see Levenshtein
        \cite{lev76}). This is
followed by using the GV bound in the Hamming space.
\begin{align*}
A_q(n,d) &\le \frac{q^n}{|SW(n,q,r)|} A_q^{SW}(n,d,r) \\
\Rightarrow \frac 1n \log_q A_q^{SW}(n,d,r) &\ge \frac1n \log_q A_q(n,d) + 
    h_q\left( 1-\frac rn\right) \\
        &\quad - 1 - o(1)\\
        &\ge h_q\left( 1-\frac rn\right) - \frac 1n \log_q
        V(\integers_q^n,d-1)\\
        &\quad    - o(1)\\
        &= h_q\left( 1-\frac rn\right) - h_q\left( \frac{d-1}n\right) - o(1),
\end{align*}
where $h_q(x)$ is the $q$-ary entropy function and $V(\integers_q^n,d-1)$ is
the volume of the ball of radius $d-1$ in the Hamming space.
Similarly, for the bounded symbol weight space, we obtain
\begin{align*}
\frac 1n \log_q  A_q^{SW}(n,d,\le{r})&\ge1 - h_q\left(\frac {d-1}n\right) -o(1).
\end{align*}
For a constant $q$, the asymptotics of these expressions are as given in
\eqref{eq:lower-bound-gv}.

For $q$ growing with $n$, 
the upper bound on the symbol weight codes is provided by the Singleton
bound,
    $$
A_q^{SW}(n,d,r)\le A_q^{SW}(n,d,\le{r})\le A_q(n,d)\le q^{n-d+1}.
    $$
Using the fact that $h_q(x) = x$  in the limit as $q\to\infty$, we get the
results as stated in the theorem.
In particular for $r=o(n)$, $\lim_n h_q(1-r/n) = 1$ and $\lim_n
h_q((d-1)/n) = \delta$. Since $A_q^{SW}(n,d,\le{r})$ is greater than
$A_q^{SW}(n,d,\ceil{n/q})$, it gives the  result stated in
(\ref{eq:aqw-rate-bd-r}).
\end{IEEEproof}
%\hfill\qed\\
\vspace{1mm}
The lower bound \eqref{eq:lower-bound-gv}
in the theorem may be interpreted as a GV-type bound in the
symbol weight space that can be obtained if 
the volume of a ball of radius $d-1$ in the
symbol weight space is upper bounded by the volume of a ball of radius
$d-1$ in the Hamming space. Since, the symbol weight space is not
ball-homogeneous, that is, the size of the balls of radius $d-1$
depends on the center, we adopt the above method to derive the GV-type lower
bound.\footnote{For certain
parameters, better lower bounds on $A_q(n,d)$, for instance from algebraic
geometry codes, can improve on this GV bound on $A_q^{SW}(n,d,r)$.}

In the following theorem we provide an improvement on the upper bound for
a {\em constant} symbol weight code with symbol weight $r$.\\
\begin{theorem}
\label{thm:large-r}
    Let $\ceil{n/q}\le r\le {2n}/{3},\ q=\theta n^\epsilon$ with
    $0\le\epsilon\le1.$ Let $d$ satisfy 
    $r\le d.$ For $n\to\infty,$ let
    $r/n \to \rho,$ and $d/n\to\delta.$ Then, for constant $q$,
\begin{multline*} 
    \limsup_{n\to\infty}\frac1n \log_q A_q^{SW}(n,d,r) \le
h_q\left(1-\frac32\rho\right) \\
        - (1-\rho) h_q\left(\frac{1-\frac32\rho}{1-\rho}\right) + 1 - \frac32 \rho.
\end{multline*}
For $q$ growing with $n$, we get
    $$
    \limsup_{n\to\infty}\frac1n \log_q A_q^{SW}(n,d,r) \le 1 - \frac32\rho.
    $$
\end{theorem}
The proof of this theorem relies on a Johnson-type upper bound, and a lemma
given below.
We follow some elements of the derivation of
the Singleton bound in Omrani and Kumar \cite{omr05}. However, our purpose
is to improve the Singleton bound by using the parameters of the constant
symbol weight space. The improvement mainly stems from the following
lemma.\\
\begin{lemma}
\label{lem:upper-q}
Let $d \ge r > {2n}/{3}.$ Then $A_q^{SW}(n,d,r)= q.$
\end{lemma}
%\Proof
\begin{IEEEproof}
We claim that if there are two codewords $\bfx, \bfy$ both with symbol
weight $r$, then the symbol which repeats $r$ times must be different in
the two codewords.
Suppose not. Then the two codewords $\bfx,\bfy$ must have at least $n - 2(n-r)$
coordinates which contain the same symbol. Thus the distance between the
codewords is at most $d \le 2(n-r)$ which implies $d<{2n}/{3}$, since
$r>{2n}/{3}$. This is a contradiction.
We get $A_q^{SW}(n,d,r)\le q.$
To show the opposite inequality, let $\bfx$ be a word with symbol weight $r,
\,     r>{2n}/{3}.$ Then $\bfx+\alpha\mathbf{1}, \ \alpha\in\integers_q$,
where $\mathbf1$ is the all-one codeword, are also codewords with
symbol weight $r.$ This establishes that $A_q^{SW}(n,d,r)\ge q$. 
%\hfill\qed
\end{IEEEproof}
\vspace{1mm}

We next give the Johnson-type bound.
\begin{lemma}
\label{lem:johnson}
$$
A_q^{SW}(n,d,r) \le \floor{ \frac{nq}{n-r} A_q^{SW}(n-1,d,r) }.
$$
\end{lemma}
%\Proof
\begin{IEEEproof}
Consider the code-matrix of the constant symbol weight code with
parameters $(n,M,d,r)_q$. Any row of the code-matrix has at least one
symbol of frequency $r.$ Fix one symbol of frequency $r$ in each
row. There
are a total $M(n-r)$ symbols in the code-matrix which  do not
contribute to the symbol weight in any codeword. The average
number of symbols, averaged over the $n$ columns, with frequency at
most $r$ is then $M(n-r)/n.$ The average number per symbol, averaged over
$n$ columns and $q$ symbols is $M(n-r)/(qn)$. 
Thus, there exists at least one symbol $\alpha$ and at least one column $m$ such that
the subcode consisting of the symbol $\alpha$ in column $m$ has size at
least $M(n-r)/(nq).$ Discarding the coordinate corresponding to $m$ gives
us the bound as stated in the Lemma. %\hfill\qed
\end{IEEEproof}
\vspace{1mm}
%\noindent\textsc{Proof of Theorem \ref{thm:large-r}:}
\begin{IEEEproof}[Proof of Theorem \ref{thm:large-r}]
We now proceed to prove the theorem. Apply Lemma \ref{lem:johnson}
recursively $l+1$ times to get
\begin{multline*}
A_q^{SW}(n,d,r) \le \Big\lfloor\frac{nq}{n-r} \cdots
    \Big\lfloor\frac{(n-l)q}{n-l-r} \times\\
        A_q^{SW}(n-l-1,d,r)\Big\rfloor \cdots \Big\rfloor.
\end{multline*}
The recursion stops for $l$ such that $r=\ceil{2(n-l-1)/3}$ and for $d\ge r$. For
this value of $r$ and $d$, $A_q^{SW}(n,d,r) = q,$ by Lemma \ref{lem:upper-q}.
The condition $l>0$ implies $r\le 2n/3$. Constraints on $l$ are obtained
from the inequalities $$2(n-l-1)/3 \le r = \ceil{2(n-l-1)/3} \le 2(n-l)/3.$$
This gives us the upper bound
\begin{align*}
A_q^{SW}(n,d,r) &\le \frac{n\cdots(n-l)}{(n-r)\cdots(n-l-r)}q^{n-3r/2+1}\\
        &= \frac{\binom n{l+1}}{\binom{n-r}{l+1}} q^{n-3r/2+1}.
\end{align*}
In the asymptotics as $n\to\infty$ we get $l/n \to 1-3/2\rho$.
This gives us the upper bounds as stated in the
theorem.% \hfill\qed\\
\end{IEEEproof}
\vspace{1mm}

In the case of  $q$ growing with $n$, this theorem
improves on the Singleton bound
$1-\delta$ for $\delta < 3\rho/2.$ 
The upper bound in Theorem \ref{thm:large-r} for constant $q$ improves on
the LP bound for certain range of parameters. The
improvements are possible only for large $\rho$ and for $q\ge5.$ For $q=2,3$
the restrictions $\delta\ge\rho$ and $\rho\le2/3$ do not leave room
for improvement. 
For constant $q$ the upper bound is in fact concave in shape.
This can be verified by taking the second derivative with respect to
$\rho$, which results in the negative expression
$ -1/(\rho(1-\rho)\ln q).$ Since this bound does not depend on $\delta$, it
seems that further improvements might be possible.

\begin{figure}[!h]
\centering
\subfigure{
\scalebox{0.44}{\includegraphics[keepaspectratio]{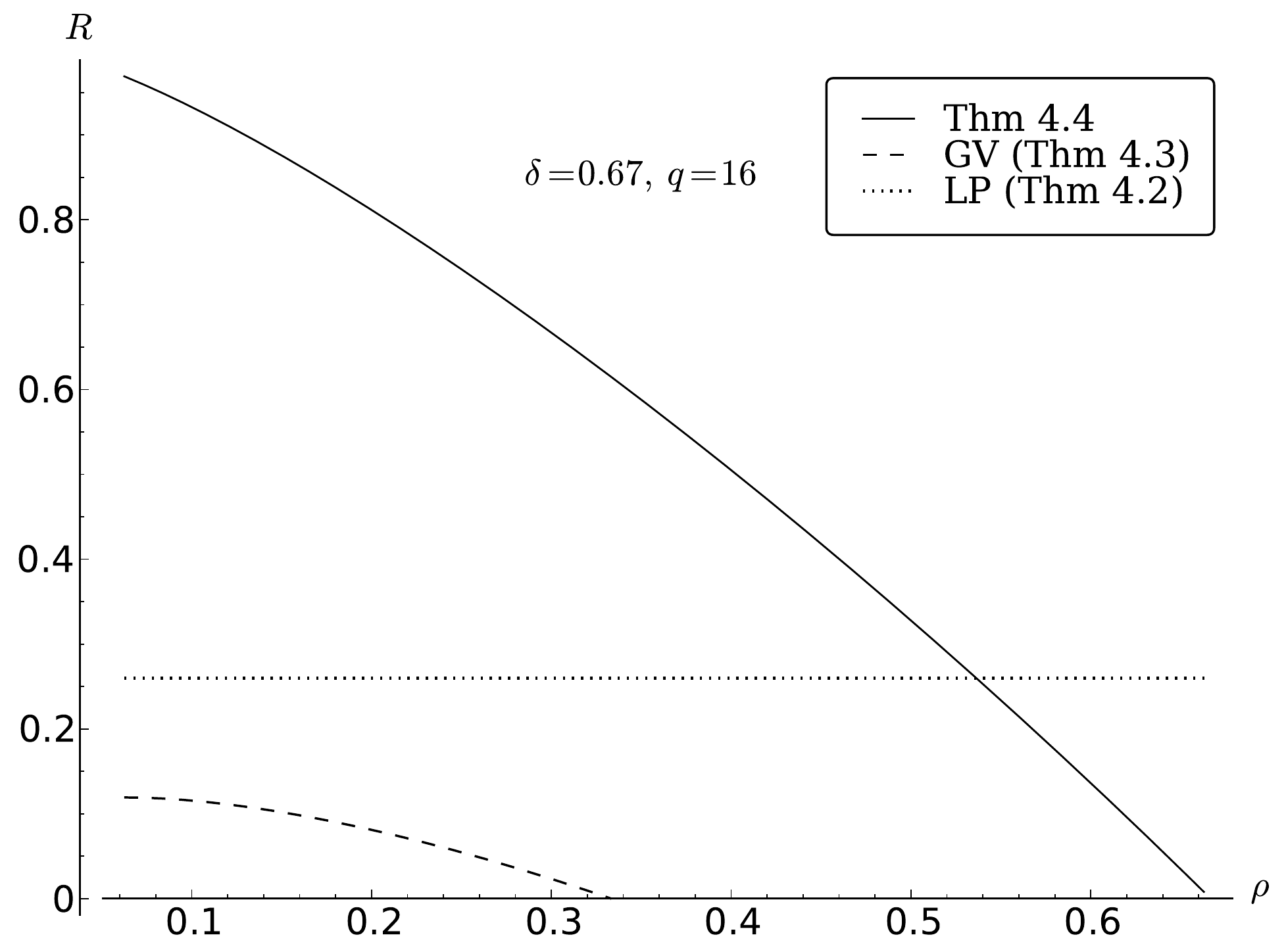}}
}
%\vspace{0.1in}
\subfigure{
\scalebox{0.44}{\includegraphics[keepaspectratio]{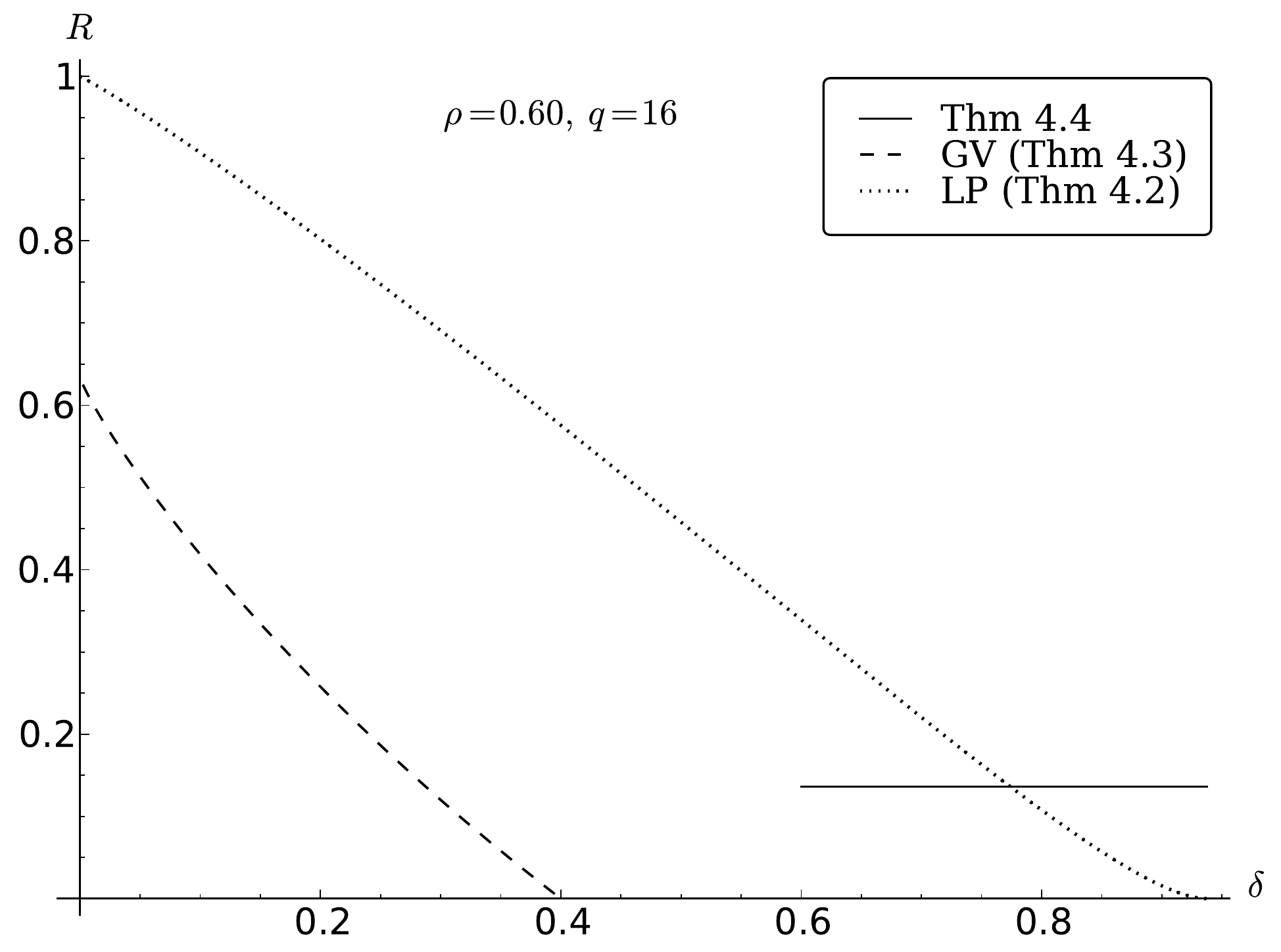}}
} 
\caption{Plots for $\delta=2/3$ and for $\rho=0.6,$ respectively, under $q=16.$}
\label{fig:constant-q}
\end{figure}

\begin{figure}[!h]
\centering
\subfigure{
\scalebox{0.44}{\includegraphics[keepaspectratio]{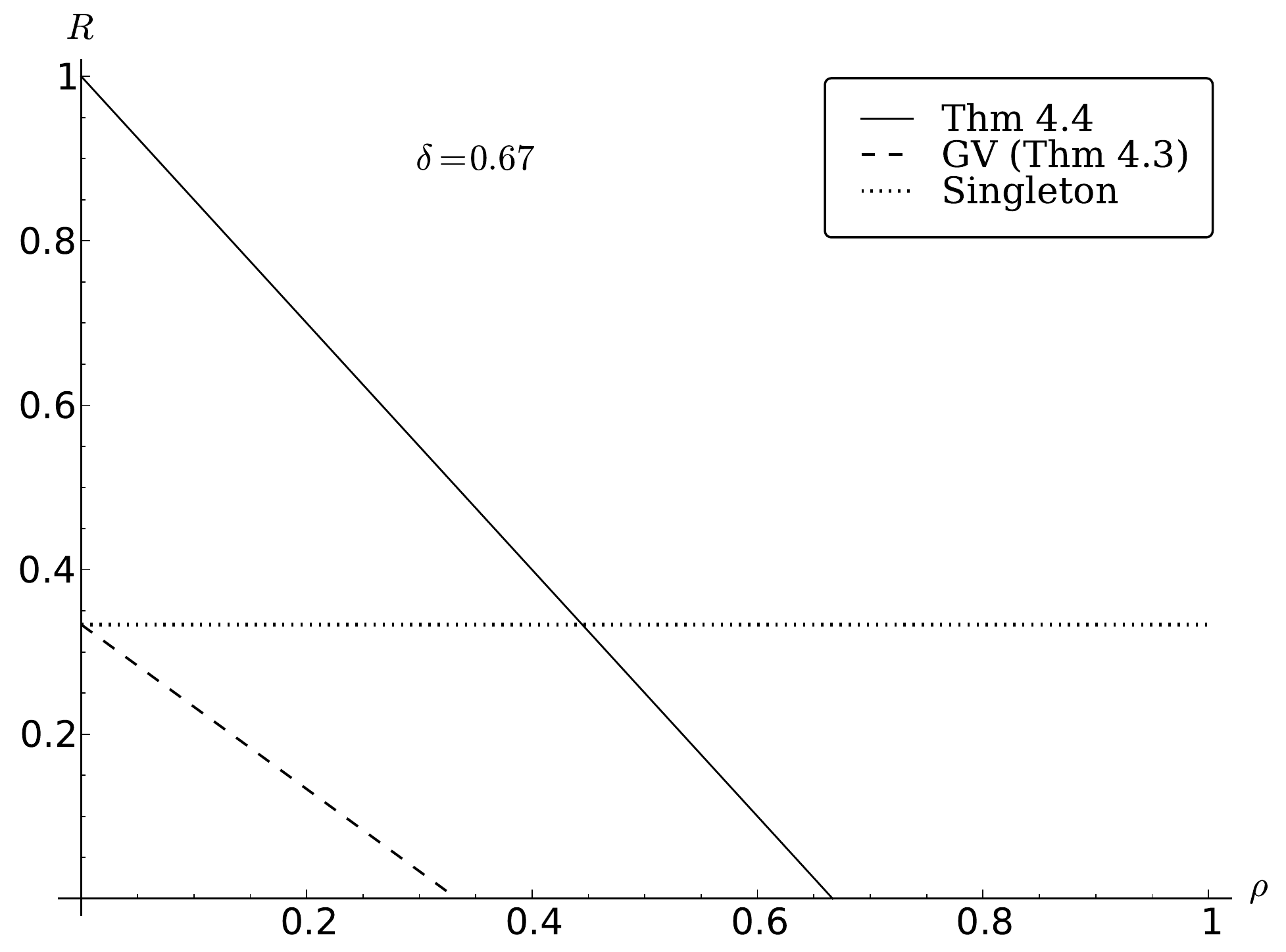}}
}
%\vspace{0.1in}
\subfigure{
\scalebox{0.44}{\includegraphics[keepaspectratio]{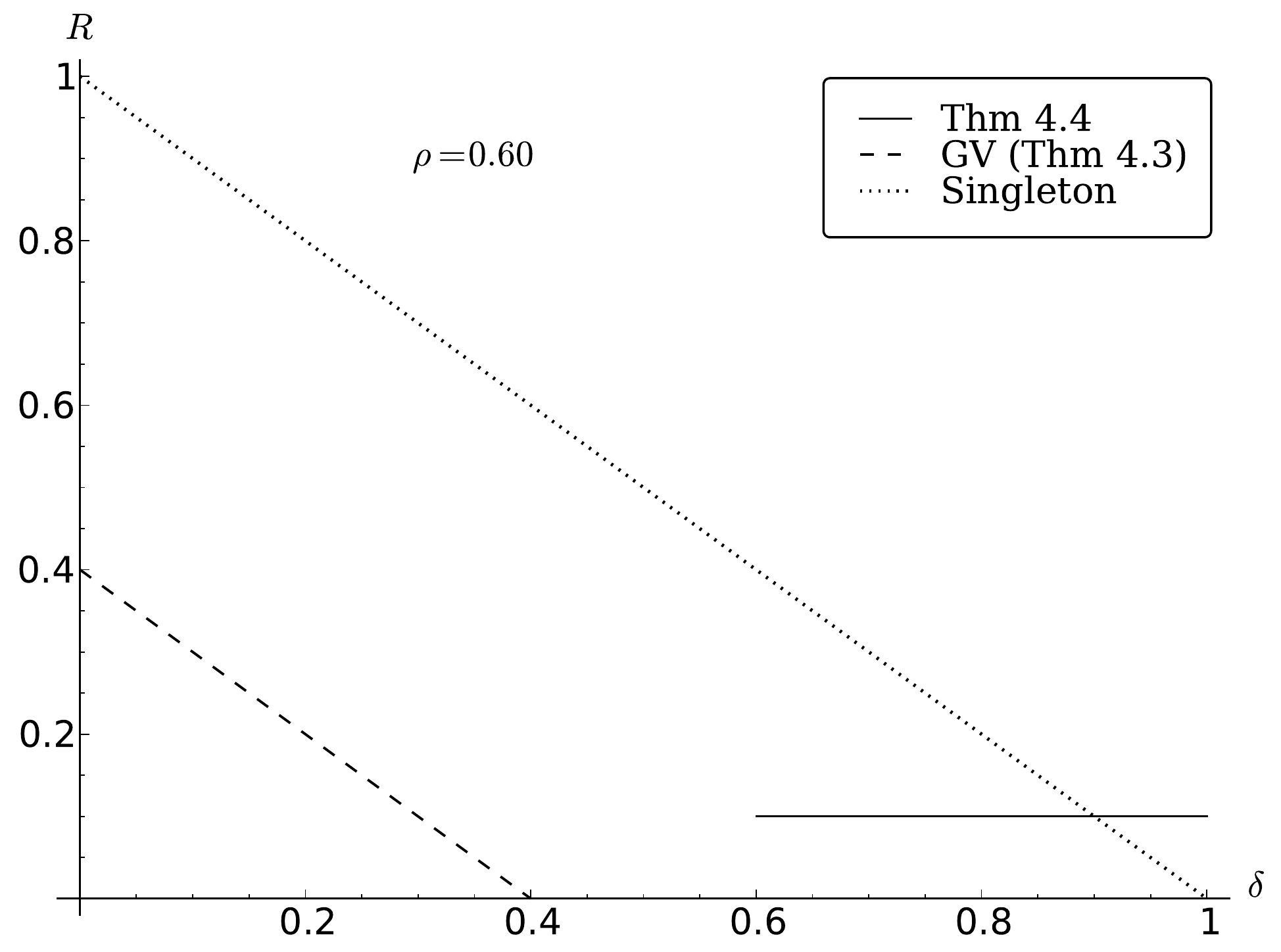}}
} 
\caption{Plots for $\delta=2/3$ and for $\rho=0.6,$ respectively.}
\label{fig:variable-q}
\end{figure}

An example plot of all the bounds are provided in figures
\ref{fig:constant-q} and \ref{fig:variable-q}. Since the improvements are
for larger $q$, we show the bounds for $q=16.$ In figure 
\ref{fig:constant-q}, the
first plot is obtained at a particular value of $\delta$ and the second
plot is obtained at a particular value of $\rho$. The improvements (over LP) are
obtained in the regions $0.536\le\rho\le0.67$ and $0.60\le\delta\le0.774$,
respectively. Figure \ref{fig:variable-q} shows the plots when $q$ is
increasing with $n$. In this case, we compare against the Singleton upper
bound. It shows improvements in the
region $\rho\le\delta\le\frac32\rho.$
Construction of codes which meet this upper bound for any
parameters is an open problem.\\[2mm]
\Remarks
\begin{enumerate}
    \item
For $q>n$ and $r=1$, Dukes \cite{duk11} provides a Singleton bound,
$
A_q^{SW}(n,d,1) \le q(q-1)\cdots(q-n+d).
$
Not surprisingly, for $q=\theta n,\, \theta > 1$ this reduces to $1-\delta$
in the asymptotics.
    \item  Missing from the list of bounds above is
a Hamming-type bound on the symbol weight codes. The lack of a simple
expression for the size of the ball is the main reason behind this omission.
\end{enumerate}
\section{Lower bound on symbol weight codes}
\label{sec:lower-bnd-sw}
As mentioned in the previous sections, the traditional means of determining
the GV-type bounds is not very useful for non-asymptotic block lengths $n.$
In this section we adopt a different approach to determine lower bounds on
the size of symbol weight codes. The lower bounds are obtained by
using corresponding constructions and lower bounds on constant composition
codes. There is a wide body of literature on CCCs
that can be used to determine these lower bounds; see \cite{che08,che10,chu04,chu06,col04, din05, din06, duk11, ges95, huc06,
huc06-2, huc10}. To the best
of our knowledge, most of the work in the literature on constant
composition codes has focused on determining constructions and bounds for either very
large distances (in the region where the Johnson bound or Plotkin bound is
applicable) or for very small distances
such as $d=2,3,4.$ For FPA and permutation codes, there do exist
constructions with distances in the ranges in between; see \cite{huc06-2,
bla79, chu06}. We note three papers in this connection. Sidorenko
\cite{sido75} provides an asymptotic upper bound on CCCs;
this is not useful for this section since we look at lower bounds.
In a very recent work, Luo and Helleseth \cite{luo11} construct CCCs of almost
uniform composition and with relative distances very close to the Plotkin
limit ($\nicefrac dn \thickapprox \nicefrac{(q-1)}q$).
The bound that we provide below requires a relatively large alphabet size
so that codes from different constant composition spaces can be combined. Hence, the
CCCs from \cite{luo11} are not useful in this context.
Chu\etal\cite{chu06} provide some lower bounds for CCCs for large
distances. We use some of the constructions from this latter work in
this section to provide examples of lower bounds on symbol weight codes.

We first describe a method to determine the size of a symbol weight code in
terms of CCCs. This method may be viewed as a generalization of an
elementary bound in Bachoc\etal\cite[Equation~(2)]{bac11} on binary bounded
weight codes to $q$-ary spaces.

Let $\bfn
= [n_0,\dots,n_{q-1}]$ and $\bfn' = [n_0',\dots,n_{q-1}']$ denote two
different compositions of $n.$ The aim here is to lower bound the size of
a symbol weight code by the sum of all possible different CCCs
which have the same symbol weight. Thus, we first need to determine the
condition on two different compositions $\bfn$ and $\bfn'$ such that any
vector $\bfc$ with composition $\bfn$ is at least distance $d$ away from
a vector $\bfc'$ with composition $\bfn'.$ 
It can be seen that $\min\{n_i,n_i'\}$ is the maximum number of coordinates
in $\bfc$ and $\bfc'$ where the $i$-th symbol is common to both. Thus,
the Hamming distance $d_H(\bfc,\bfc')$ satisfies
$$
d_H(\bfc,\bfc') \ge n - \sum_{i=0}^{q-1} \min\{n_i,n_i'\}.
$$
A sufficient condition for $d_H(\bfc,\bfc') \ge d$ to hold is
\begin{equation}
\label{eq:comp}
n - \sum_{i=0}^{q-1} \min\{n_i,n_i'\} \ge d.
\end{equation}
Let 
\begin{equation}
\label{eq:d+}
d_+(\bfn,\bfn')
\triangleq n - \sum_{i=0}^{q-1} \min\{n_i,n_i'\}.
\end{equation}
Then we obtain\\
\begin{lemma}
$d_+(\cdot,\cdot)$ is a distance function on $\cN.$
\end{lemma}
%\Proof
\begin{IEEEproof}
$d_+(\bfn,\bfn')$ is clearly symmetric. To show the triangle inequality, we
note that we can rewrite
\begin{align*}
d_+(\bfn,\bfn') &= \sum_{i=0}^{q-1} n_i - \min\{n_i,n_i'\}\\
        &=  \sum_{i} (n_i - n_i')^+,
\end{align*}
where $(x)^+ \triangleq \max\{x,0\}$. Also, for any non-negative real numbers
$x,y,z$, it can be readily verified that 
$$
(x-y)^+ + (y-z)^+ \ge (x-z)^+.
$$
Thus, we get 
\begin{align*}
d_+(\bfn,\bfn') + d_+(\bfn',\bfn'') &= \sum_i (n_i - n_i')^+ + \sum_i (n_i'
        - n_i'')^+\\
        &= \sum_i (n_i - n_i')^+ + (n_i'-n_i'')^+ \\
        &\ge \sum_i (n_i  - n_i'')^+\\
        &= d_+(\bfn,\bfn'').
\end{align*}
Since $(n_i - n_i')^+ \ge 0$, we get that $d_+(\bfn,\bfn') = 0$ if and only
if $n_i = n_i'$ for all
$i=0,\dots,q-1.$% \hfill\qed\\
\end{IEEEproof}
\vspace{2mm}

Let $\cN(r,d) \subset \cN(r)$ (resp. $\cN(\le{r},d) \subset \cN(\le{r})$) 
be such that for any distinct
$\bfn,\bfn'\in\cN(r,d)$ (resp. $\cN(\le{r},d)$) 
we have $d_+(\bfn,\bfn')\ge d.$
We can now readily give a lower bound on the size of symbol weight codes in
terms of the CCCs:
\begin{equation}
\begin{split}
A_q^{SW}(n,d,r) \ge \sum_{\bfn\in\cN(r,d)} A_q(\bfn,d),\\
A_q^{SW}(n,d,\le{r}) \ge \sum_{\bfn\in\cN(\le{r},d)} A_q(\bfn,d).\\
\end{split}
\label{eq:sw-cc}
\end{equation}
For large $n$ and $q$ the size of the set $\cN(r)$ becomes very large.
Hence finding all
the compositions in $\cN(r)$ which are separated by distance at least $d$
is difficult. We instead seek lower bounds on $\cN(r,d)$ so that the size
of the symbol weight codes can be more easily expressed in terms of the
sizes of either one or a few CCCs.

\Remark
Note that $d_+(\cdot,\cdot)$ is a metric on a ``simplex'' which
intersects each axis at (Euclidean) distance $n$ from the origin. In
particular, the components of $\bfn$ need not be restricted to integers for
$d_+(\cdot,\cdot)$ to become a metric. Also, $n$ need not be restricted
to be an integer.

\subsection{Lower bounds on $|\cN(r,d)|$}
\label{sec:lower-bd-nrd}
In this section we determine lower bounds to the size of $\cN(r,d)$.
To get these lower bounds, we first obtain a relation between the Hamming
distance between two compositions and the distance between two compositions
as given by (\ref{eq:d+}).\\
\begin{lemma}
\label{lem:n-nprime-dH}
For any two compositions $\bfn,\bfn'\in \cN$, if $d_H(\bfn,\bfn') = 2d$ then
$d_+(\bfn,\bfn')\ge d$.
\end{lemma}
\vspace{1mm}
\Proof
Define two sets $I^+=\{i: n_i > n'_i\}$ and $I^-=\{i:n_i < n'_i\}$.
Clearly, in the rest of the coordinates, $n_i = n'_i$.  Then we get the
following set of equalities.
%\begin{align*}
%\sum_{i=0}^{q-1} n_i &= \sum_{i=0}^{q-1} n'_i\\
%\Leftrightarrow \qquad\sum_{i\in I^+} n_i + \sum_{i\in I^-} n_i &= \sum_{i\in I^+}
%    n'_i + \sum_{i\in I^-} n'_i \\
%\Leftrightarrow \qquad\ \quad\sum_{i\in I^+} (n_i - n'_i) &= \sum_{i\in I^-} (n'_i
%        - n_i)\\
%\Leftrightarrow \quad\,\sum_{i\in I^+\cup I^-} (n_i - n'_i)^+ &= \sum_{i\in I^+
%    \cup I^-} (n'_i - n_i)^+
%\end{align*}
\begin{IEEEeqnarray}{crl}
& \sum_{i=0}^{q-1} n_i\ &= \sum_{i=0}^{q-1} n'_i \nonumber\\
\Leftrightarrow & \sum_{i\in I^+} n_i + \sum_{i\in I^-} n_i\ &= \sum_{i\in I^+}
    n'_i + \sum_{i\in I^-} n'_i \nonumber \\
\Leftrightarrow & \sum_{i\in I^+} (n_i - n'_i)\ &= \sum_{i\in I^-} (n'_i
        - n_i) \nonumber\\
\Leftrightarrow & \sum_{i\in I^+\cup I^-} (n_i - n'_i)^+\ &= \sum_{i\in I^+
    \cup I^-} (n'_i - n_i)^+.
\label{eq:n-nprime}
\end{IEEEeqnarray}
Note that the LHS and RHS of the last equation are
both equal to $d_+(\bfn,\bfn')$. Let $|I^+| = x$, then since
$d_H(\bfn,\bfn') = 2d$, we get $|I^-| = 2d-x.$ Using the fact that the
difference $n_i - n'_i\ge 1$, for $i\in I^+$ and $n'_i - n_i\ge 1$ for $i
\in I^-$, we get
\begin{align*}
d_+(\bfn,\bfn') &\ge \max\{x, 2d-x\}\\
        &\ge \min_{1\le x\le 2d-1} \max\{x,2d-x\}\\
        &= d.
    \qquad\qquad\qquad\qquad\qquad\qquad\qquad\qquad\qquad{\qed}
\end{align*}
This lemma immediately allows us to use existing GV bounds in various
spaces (under Hamming distance) to derive lower bounds on $|\cN(r,d)|.$
\subsubsection{Lower bound from a permutation code on $S_{r}$}
Let the alphabet set be $\{0,\dots,r-1\}$, that is, $q=r.$ In every word
of length $n=r(r+1)/2$, let all the $r$ symbols occur such that the frequencies
of the symbols are in the set $\{1,\dots,r\}$ and all the frequencies
occur.
Because of this construction, given $r$ the values of $n,q$ are restricted
as given above.
Using the GV lower
bound on the permutation code with Hamming distance at least  $2d$ between two
codewords gives us the lower bound
\begin{equation}
\label{eq:sw-perm}
|\cN(r,d)| \ge \frac{r!}{V(2d-1,S_{r})},
\end{equation}
where $V(2d-1,S_{r})$ is the volume of the ball of radius $2d-1$ in
$S_{r}$.% Note that $q$ and $r$ both
%grow as $\sqrt{n}$ for large $n$.
%
%
%
%
%
\subsubsection{Lower bound for general $q$}
\label{sec:lower-bd-nrd-qn}
%Let $q=\theta n$, where $0< \theta\le 1 $ is a constant.
As explained in the proof of Theorem \ref{thm:aqw-rate-q-n-r-const}, a lower
bound on constant symbol weight codes is provided by an Elias-type bound in the
Hamming space.
We can obtain
another lower bound on constant symbol weight codes by considering lower
bounds on $\cN(r,d)$. A lower bound 
on $\cN(r,d)$ is obtained by letting $k$ of the symbols $\{0,\dots,q-1\}$
repeat $r$ times in every codeword of the constant symbol weight code and the
remaining $q-k$ symbols satisfy the condition that there are 
$(q-k)\{\frac{n-rk}{q-k}\}$ symbols with composition $l_1
= \ceil{\frac{n-rk}{q-k}}$ and $(q-k)(1-\{\frac{n-rk}{q-k}\})$ symbols
with composition $l_0 = \lfloor{\frac{n-rk}{q-k}}\rfloor$.
Denote this composition by $\bfn = \bfn(l_0,l_1,k,r)$, that is,
$$
\bfn(l_0,l_1,k,r) = [\underbrace{r,\ldots, r}_{k},
\underbrace{l_0,\ldots ,l_0}_{(q-k)\{\frac{n-rk}{q-k}\}},
\underbrace{l_1,\ldots ,l_1}_{(q-k)(1-\{\frac{n-rk}{q-k}\})}].
$$
Because of
the above choice of the repetitions of each symbol we seek a ``binary
constant weight code'' in $\cN(r,d)$ where $q-k$ coordinates have the value
$l_0$ or $l_1$ and the rest $k$ coordinates have the value $r.$ We also
want the Hamming distance between distinct codewords to be at least $2d.$
Denote the maximum size of a binary constant weight code of length $n$,
weight $w$ and minimum distance $d$ by $A_2(n,d,w)$. 

The GV bound under the above constraints is
$$
|\cN(r,d)| \ge A_2(q,2d,k) 
    \ge \frac{\binom qk}{\sum_{i=0}^{2d-2} \binom ki \binom {q-k}i}.
$$
Note that we get $2d-2$ in the denominator (instead of $2d-1$) since
the binary constant weight space affords only even distances.
From Levenshtein \cite{lev70} we know that the lower bound is
significant and grows exponentially as $2^{q T}$, for some constant $T
\equiv T(q,k,d)$ only when the following conditions are satisfied
\begin{align*}
\frac q2\left(1 - \sqrt{1 - \frac {4d}q}\right) &\le k \le
    \frac q2\left(1 + \sqrt{1 - \frac {4d}q}\right), \\
d &\le k\left(1 -\frac kq\right).
\end{align*}
The above lower bounds  on $|\cN(r,d)|$ give lower bounds on
symbol weight codes as follows.
\subsection{Lower bounds on codes}
The lower bound on $A_q^{SW}(n, d, r)$ can be stated as follows.
\begin{theorem}
\label{thm:finite-case-csw}
We get the following results for different compositions.
\begin{enumerate}
    \item 
For $\bfn = [1,\dots,r]$, we get
$$
A_q^{SW}(n,d,r) \ge
    \frac {r!}{V(2d-1,S_{r})} A_q(\bfn,d).
$$
    \item 
For $\bfn = \bfn(l_0,l_1,k,r)$, we get
\begin{align*}
A_q^{SW}(n,d,r) & \ge A_2(q, 2d, k) A_q(\bfn, d)\\
    & \ge \frac {\binom qk}{\sum_{i=0}^{2d-2} \binom ki \binom {q-k}i} 
    A_q(\bfn,d).
\end{align*}
    \item 
Let $k_1\ge k_0 = \max\{n-(r-1)q,1\}$, and $b\equiv b(k_1)
= \floor{\frac{\lfloor n/r\rfloor - k_1}{2d}}$. Then
\begin{multline}
\label{eq:size-constant-finite-bachoc}
A_q^{SW}(n,d,r) \ge \max_{k_0\le k_1\le \floor{n/r}}
        \sum_{i=0}^b A_2(q,2d,k_1+2di)\times \\
                A_q(\bfn(l_0,l_1,k_1+2di,r), d).
\end{multline}
\end{enumerate}
\end{theorem}
%\Proof
\begin{IEEEproof}
The first two results follow immediately from the lower bounds on
$|\cN(r,d)|$. In part 1, each codeword in the permutation code corresponds
to a rearrangement of the composition $\bfn=[1,\dots,r].$ In part 2, each
codeword in the binary constant weight code corresponds to a rearrangement
of the composition in $\bfn = \bfn(l_0,l_1,k,r)$.

For the third result, we include a larger range of CCCs.
The expression is obtained by taking constant
composition codes from separate constant composition spaces
$\bfn(l_0,l_1,k_1+2di,r)$, whose compositions are separated by a Hamming
distance of at least $2d$. Two different compositions
$\bfn(l_0,l_1,k_1+2di,r)$ and $\bfn(l_0,l_1,k_1+2d(i+1),r)$
correspond to taking binary constant weight codes with weights separated
by $2d.$ Note that this choice of separate compositions
corresponds to a binary bounded weight code, as studied in \cite{bac11}.
\end{IEEEproof}
%\hfill\qed
\vspace{1mm}

There is a trade-off between the size of the constant composition
space with composition $\bfn(l_0,l_1,k,r)$ and the size of the constant
weight code in $\integers_2^q$. The size of the constant weight code in
$\integers_2^q$
is substantial only for large $k$ around $q/2.$ On the other hand, the size
of the constant composition space is large for small $k$, thus potentially
allowing for a larger CCC.

The lower bound in equation \eqref{eq:size-constant-finite-bachoc} is in
fact useful in the case of a bounded symbol weight code with symbol weight
at most $r$.  It is unclear how to combine codes of different symbol
weights $s$, where $\ceil{n/q}\le s\le r$,
such that we can obtain a computable expression.
We instead use equation \eqref{eq:size-constant-finite-bachoc} and optimize
over the different symbol weights $s$ and the smallest weight $k_1$. Note
that for a given symbol weight $s,$ the quantity $k_1$ corresponds to the
minimum number of symbols with frequency $s$ that we include in our
estimate.\\
\begin{theorem}
\label{thm:finite-case-sw}
\begin{align*}
        A_q^{SW}(n,d,\le{r}) & \ge \max_{\ceil{\frac nq}\le s\le r}A_q^{SW}(n,d,s)\\
                             & \ge \max_{\ceil{\frac nq}\le s\le r} \max_{
                                k_1\le \floor{\frac ns}}
        \sum_{i=0}^{b(s)} A_2(q,2d,k(i,s)) \\
   & \qquad \times A_q\big(\bfn(l_0(i,s),l_1(i,s),k(i,s),s),d\big),
\end{align*}
\end{theorem}
where $k_1\equiv k_1(s)\ge \max\{n-(s-1)q,1\},\  
k \equiv k(i,s) = k_1(s) + 2di,\ 
b(s) = \floor{\frac{\lfloor n/s\rfloor - k_1}{2d}},\ 
l_0(i,s) = \lfloor\frac{n-sk}{q-s}\rfloor,$ and $
l_1(i,s) = \ceil{\frac{n-sk}{q-s}}$.

\subsection{Numerical examples} % (fold)
\label{sub:examples}
We consider some numerical examples in order to show how the expressions in the
previous section can be used to obtain lower bounds on symbol weight codes.
We adopt the exponential notation of Chu\etal\cite{chu06} to denote a
composition in a compact form. The notation $n_0^{t_0}n_1^{t_1}\dots
n_{q-1}^{t_{q-1}}$ is used to denote the composition
$$
[\underbrace{n_0,\dots,n_0}_{t_0},\underbrace{n_1,\dots,n_1}_{t_1},\dots,\underbrace{n_{q-1},\dots,n_{q-1}}_{t_{q-1}}].
$$
We also recall the notion of a refinement of a composition from the same
work. The composition $\bfn = [n_0,\dots,n_{q-1}]$ is called a \emph{refinement}
of a composition $\bfm = [m_0,\dots,m_{p-1}]$ if there is a partition
$I_0,\dots,I_{p-1}$ of $\{0,\dots,q-1\}$ such that $\sum_{i\in I_j} n_i
= m_j,$ for every $j.$ We write 
$\bfn \preccurlyeq \bfm$ if $\bfn$ is a refinement of $\bfm.$
This notion is important because of the following
inequality (see \cite{chu06}):
\begin{equation}
\label{eq:refinement-bound}
A_q([n_0,\dots,n_{q-1}], d)  \ge A_p([m_0,\dots,m_{p-1}], d).
\end{equation}
Below, we use lower bounds on FPAs, where the lower bound is taken from
\cite{chu06}. Using equation \eqref{eq:refinement-bound}, lower bounds on
CCCs are obtained from the lower bounds on the FPAs.
The lower bound on FPA mentioned below rely on the existence of certain
(generalized) distance preserving mappings from $\integers_q^n$ to the
permutation space $S_n$ (see \cite{chu06}). The distances between
compositions used in this section are all taken in the $d_+(\cdot,\cdot)$
metric, unless mentioned otherwise.\\
\begin{example}
\label{ex:1}
In this example we show how Theorem \ref{thm:finite-case-csw} and equation
\eqref{eq:sw-cc} can be used.
    We know from \cite[Example 3.7]{chu06} that $A_4(6^4, 7)\ge 2^{12}$.
Since $1^45^4 \preccurlyeq 6^4,$ we immediately obtain that $A_8(1^45^4,
7)\ge 2^{12}$. In this case, $q = 8$ and $d = 7$ and the number of
symbols occurring with frequency $5$ is $k=4$. Hence Theorem
\ref{thm:finite-case-csw} is not applicable. But equation
\eqref{eq:sw-cc} can be applied directly. For instance the compositions
$1^45^4$ and $5^41^4$ satisfy $d_+(1^45^4,5^41^4) = 4(1-5)^+ + 4(5-1)^+
= 16$ which is greater than 7. Thus, $A_8^{SW}(24,7,5)\ge 2\cdot 2^{12}.$
\end{example}

In fact, the compositions $\bfn=1^45^4$ and $\bfn' = 5^41^4$ have the special property
that if a symbol $i$ has different compositions $n_i, n_i'$ then $|n_i
- n_i'| = 4.$ We can exploit this property to get a variant of Lemma
\ref{lem:n-nprime-dH} below.\\
\begin{lemma}
\label{lem:n-nprime-dH-special}
For two compositions $\bfn,\bfn'\in\cN$ let $d_H(\bfn, \bfn') = D.$
For $i=0,\dots,q-1$, if either
$|n_i - n_i'|\ge a > 0$ or it is zero, then 
$d_+(\bfn, \bfn')\ge Da/2.$
\end{lemma}
\Proof
The proof is very similar to the proof of Lemma~\ref{lem:n-nprime-dH}.
Let $I^+, I^-$ be as defined in the proof of that lemma, and let 
$|I^+| = x,\ |I^-| = D-x.$ Finally, use equation \eqref{eq:n-nprime}
to get
\begin{align*}
    d_+(\bfn, \bfn') &\ge a \max\{x, D-x\}\\
            &\ge a \min_{1\le x \le D-1}\max\{x, D-x\}\ge Da/2.
\qquad\qquad\qed
\end{align*}

\textit{Example~\ref{ex:1}}~(contd.): Returning to this example, we know
that if two compositions differ in a symbol, then the difference is 4, and
so $a=4.$ We will have the distance between two compositions at least 7 if
we ensure (using Lemma \ref{lem:n-nprime-dH-special}) that $Da/2 \ge 7$,
that is, $D \ge \ceil{14/4} = 4.$ Using $q=8, D=4, k=4$, we get that
 $A_8^{SW}(24,7,5)\ge A_2(8,4,4)
A_8(1^45^4,7)$. The size of the binary code is obtained from
Agrell\etal\cite{agr00}: $A_2(8,4,4) = 14.$ This gives the much improved
lower bound $A_8^{SW}(24,7,5)\ge 14\cdot 2^{12}.$\\

\begin{example}
\label{ex:12}
We continue with the previous example and a finer refinement
$1^82^8 \preccurlyeq 6^4.$ 
We show in this example that directly using equation \eqref{eq:sw-cc} can
lead to a better bound, compared to Theorem~\ref{thm:finite-case-csw}.
In this case, we apply
Theorem~\ref{thm:finite-case-csw} with $k=8,$ $2d = 14,$ $q=16,$ and $
\bfn(l_0,l_1,k,r) = 1^82^8.$ From the table of constant weight codes in
\cite{agr00}, we get $A_2(16,14,8) = 2.$ This gives
$
A_{16}^{SW}(24,7,2) \ge 2\cdot 2^{12}.
$
However, this bound can be improved by using equation \eqref{eq:sw-cc}.
A greedy search through the compositions with the maximum value of each
part being 2, show that the
compositions $0^31^22^{11},$ $ 1^12^40^31^1 2^7,$ $  1^12^80^31^12^3,$ $
2^11^12^31^12^31^12^30^21^1,$ and $2^30^12^40^12^30^12^20^1$ are mutually at distance at least
7 from one another. Since each of them is a refinement of $6^4$ we get that
$
A_{16}^{SW}(24,7,2)\ge 5\cdot 2^{12}$.\\
\end{example}
Finally, we look at an example which considers the bounded symbol weight
and demonstrate the use of Theorem~\ref{thm:finite-case-sw}.
\begin{example}
Consider the refinement $3^8 \preccurlyeq 6^4.$ We can consider a code in
this space to be embedded in the constant composition space with 16
symbols. Thus, we get $A_{16}(3^80^8, 7)\ge2^{12}$.
We use the fact that if we consider all compositions containing 8 symbols
occurring with frequency $3$, then the difference of frequency between two
symbols from different compositions is either 0 or 3. Thus, $a = 3$, $k
= 8,$ and $q=16.$ Using Lemma \ref{lem:n-nprime-dH-special},
we need to ensure that $Da/2\ge 7\Rightarrow D\ge 5.$
We get the lower bound $A_{16}(24, 7, 3) \ge A_2(16,5,8)
A_{16}(3^80^8, 7)$. From \cite{agr00} we have $A_2(16,5,8) = A_2(16,6,8)
\ge 120.$
%We use the fact that
%the compositions $3^80^8,$ $ 0^83^8,$ $ 3^20^43^40^4,$ $ 0^43^40^43^4,
%3^20^23^20^23^20^23^20^2,$ $0^23^20^23^20^23^20^23^2,$
%$ [3,0,3,0,3,0,3,0,3,0,3,0,3,0,3,0],
%$ $
%[0,3,0,3,0,3,0,3,0,3,0,3,0,3,0,3]$ are all separated by a distance of at least 9, and hence
%by using equation \eqref{eq:sw-cc}, we get
Hence, 
$
A_{16}^{SW}(24,7,\le{3}) \ge \max\{A_{16}^{SW}(24,7,3),
A_{16}^{SW}(24,7,2)\} \ge 120\cdot 2^{12}.
$ In this particular example, using the lower bound in equation
\eqref{eq:size-constant-finite-bachoc} on the constant symbol weight codes
with $k_1 = 1$
does not yield a better bound, primarily due to the absence of a known good
lower bound on the corresponding CCC. The improvement mainly stems from the
fact
that  we use
a large binary constant weight code with weight $k_1 = q/2$.\\
\end{example}

As is evident from the above examples, Theorems \ref{thm:finite-case-csw}
and \ref{thm:finite-case-sw} help in actual computation of the bounds.
%because we rely on known results on sizes of codes in the binary Hamming
%space and constant composition space.
In Example \ref{ex:12}, the number of
ordered partitions of $n=24,$ into $q=16$ parts with each part
taking values between 0 and 2, inclusive, is $|P(24,16,2)| = 258570.$ A 
non-exhaustive greedy search could only find
5 compositions. It is computationally difficult to search exhaustively in
such a large space. Similarly, for larger lengths and alphabet sizes,
finding the size of, and compositions in, $\cN(r,d)$ is difficult. 
Instead, by relying on known bounds on binary constant weight codes and
CCCs, 
we compute the sizes of the symbol weight codes more
easily.

% subsection Numerical examples (end)

\section{Constructions of symbol weight codes} % (fold)
\label{sec:constructions}
In this section we determine constructions of  symbol weight
codes. Versfeld\etal\cite{ver05, ver10} provided constructions of
bounded symbol weight codes from Reed-Solomon codes. We seek to obtain  constant symbol weight
codes with positive rate and positive relative distance.
The following two constructions provide us with such codes with positive
rate and positive relative distance, given that we already have
a constant symbol weight code with positive rate and positive relative
distance.
%Recall that an FPA is a CCC with composition $\bfn
%= [n/q,\dots,n/q]$ such that $q$ divides $n.$

\textbf{$\bfu|\bfv$ construction:} Let $\cC$ be a constant symbol weight code with
    parameters $\cC(n,M, d, r)_q$ over $\integers_q$. Let
    $\cC'$ be an FPA over $\integers_q$ with parameters
    $\cC'(r'q, M', d')$. Then the
    $\bfu|\bfv$ construction results in a code $\mathcal{D}$. It is obtained by
    taking all codewords as follows:
    $$
    \mathcal{D} = \{ (\bfu,\bfv): \bfu\in\cC, \bfv\in\cC'\}.
    $$
    The code $\mathcal{D}$ has parameters $\mathcal{D}(n+r'q, MM',
            \min\{d,d'\}, r+r')_q$ over $\integers_q$. In particular if the code
    $\cC$ had the minimum symbol weight $r = \ceil{n/q}$ then so does the
    code $\mathcal{D}$, that is, $r+r' = \ceil{(n+r'q)/q}$.

\textbf{Concatenated construction:} Let $\cC$ be a code over
    $\integers_q$ with parameters $\cC(n, M, d)_q$. Let $\cC'$ be an FPA with
    parameters $\cC'(rp, M', d')_p$ over $\integers_p$, such that $M'\ge q.$
    The concatenated code
    $\mathcal{D}$ with $\cC'$ as the inner code and $\cC$ as the outer code
    has parameters $\mathcal{D}(nrp, M, dd', rn)_p$. It is obtained by
    replacing every $q$-ary symbol of $\cC$ with a codeword from $\cC'$. 
    In particular, the resulting code $\mathcal{D}$ has the minimum symbol
    weight $npr/p = nr.$

\subsection{Constructions from Reed-Solomon codes}
In this section, we use the symbol $k$ to denote the dimension of the
Reed-Solomon code.
Let $\cc[n,k,d]_q$ be a Reed-Solomon code over a finite field 
$\ff_q$ with $d=n-k+1$ and
$n=q-1.$ Versfeld\etal\cite{ver05, ver10} showed that aside
from the Reed-Solomon codewords which correspond to a constant polynomial,
the Reed-Solomon code has maximum symbol weight of $n-d = k-1.$ It is also
established in the same works that there exists a coset of
the Reed-Solomon code such that the maximum symbol weight of any codeword
in the code is at most $n-d+1.$ These codes belong to the bounded symbol weight space
$SW(n,q,\le{r})$. By the Singleton bound, these are optimal codes.

In this section we establish several results which show that subsets of
Reed-Solomon codes or their cosets can achieve the GV-type lower bound in
Theorem~\ref{thm:aqw-rate-q-n-r-const}. First, Lemma~\ref{lem:rs-coset}
below shows that for any \emph{constant} symbol weight $r$, there exists
a coset of the Reed-Solomon code that attains the GV bound asymptotically.
In Theorem~\ref{thm:rs-code}, we provide a more explicit description of a
subset of the Reed-Solomon code itself that has the {constant} symbol
weight $r$ for $r\ge n/2$, such that it attains the GV bound
asymptotically.  Since $k-1\ge r$, this also means that this latter result
holds only for Reed-Solomon codes with rate more than $1/2.$
%Asymptotically, this
%subset of the Reed-Solomon code attains the lower bound derived in Theorem
%\ref{thm:aqw-rate-q-n-r-const}. 

We first show by an averaging argument that the GV bound can be achieved by
subcodes of cosets of Reed-Solomon codes.
\begin{lemma}
\label{lem:rs-coset}
Let $\cC[n,k,d]_q$ be a family of Reed-Solomon codes. For $n\to\infty$ let
$r/n\to\rho,$ and $d/n\to\delta.$ Then there exists a family of
subcodes $\cC'$ which is a subset of some coset of the code $\cC$ such that
$$
\liminf_{n\to\infty} \frac1n \log_q |\cc'| \ge 1-\rho-\delta.
$$
\end{lemma}
%\Proof
\begin{IEEEproof}
Let $\cc_1,\dots,\cc_{q^{n-k}}$ denote the cosets of the Reed-Solomon code.
Since the cosets of the Reed-Solomon code are disjoint and they partition
the Hamming space, we have
$$
\sum_{i=1}^{q^{n-k}}|\cc_i\cap SW(n,q,r)| = |SW(n,q,r)|.
$$
Thus, the average size of the intersection of a coset with the space
$SW(n,q,r)$ is $|SW(n,q,r)|/q^{n-k}$. Hence, there exists at least one
coset whose intersection with $SW(n,q,r)$ has size at least this average.
In the asymptotics for $n,q\to\infty$, we get the result stated in the
Lemma.%\hfill\qed
\end{IEEEproof}
\vspace{1mm}

In the remaining part of this section, we give a more explicit description
of a subcode of the Reed-Solomon code with rate equal to the GV-type bound.
The derivation of this result uses a lemma and a proposition
stated below. The Proposition \ref{prop:sw-le-wt-distr} below states that
asymptotically the rate of the constant symbol weight code with symbol
weight $r$ that is a subset of the Reed-Solomon code, can not exceed the
rate of the subcode formed by all the codewords
of weight $n-r.$ Lemma \ref{lem:wt-distr} gives an upper bound on the size
of the number of codewords of weight $n-r.$ The combination of this
proposition and the lemma imply that the rate of the constant symbol
weight code,  which is a subset of the Reed-Solomon code,
with symbol weight $r = \rho n$, can not exceed the GV-type lower
bound $1-\rho - \delta$ that we obtained in Theorem
\ref{thm:aqw-rate-q-n-r-const}, \emph{for any} $\rho,\, 0 < \rho < 1.$
Theorem \ref{thm:rs-code} below shows that this rate can be attained for
any $\rho$ satisfying $1/2 \le \rho < 1.$
We state the proposition and the lemma first, and defer their proofs 
to after the proof of the theorem.\\
\begin{proposition}
\label{prop:sw-le-wt-distr}
Let $\cc[n,k,d]_q$ be a family of Reed-Solomon codes with parameters
$n=q-1, d=n-k+1$. Let $S(r)$ denote the set of vectors with symbol weight
exactly $r,$ for $1\le r\le k-1$, and let $B_{n-r}$ denote the number of vectors of weight
$n-r.$ Then,
$$
|S(r)| \le q(q-1) B_{n-r}.
$$
\end{proposition}

\begin{lemma}
\label{lem:wt-distr}
The weight distribution $\{B_w: w=d,...,n\}$ of a linear 
Maximum Distance Separable (MDS) code
with parameters $[n,k,d]_q$
satisfies:
\begin{align*}
B_{n-r} \le \binom n{n-r}(q^{k-r}-1),
\end{align*}
for $0\le r \le k-1$.\\
\end{lemma}

The main theorem in this section is now stated below.
\begin{theorem}
\label{thm:rs-code}
    Let $\cC[n,k,d]_q$ denote the family of Reed-Solomon codes with $n=q-1$
    and $d=n-k+1$. Let $k-1\ge r\ge n/2$. For $n\to\infty$, let $r/n \to \rho$
    and $d/n\to\delta.$ There exists a family of subcodes $\cC'$ of $\cC$
    of symbol weight exactly $r$
    such that
    $$
    \lim_{n\to\infty}\frac1n \log_q |\cC'| = 1 - \rho - \delta.
    $$
\end{theorem}
\Proof
Every codeword of the Reed-Solomon code consists of coordinates which are
the evaluations at all the non-zero points of $\ff_q$, of a polynomial of
degree at most $k-1$. Let $f(x) = f_0 + f_1 x + \cdots + f_{k-1} x^{k-1}$
be a polynomial in $\ff_q.$ Let $\ff_q^* = \ff_q\setminus\{0\}.$
If $f(x)$ has symbol weight $r$ then it implies that $f(x) = \alpha$ for
some $\alpha\in\ff_q$ and for $r$ different values of $x$ in $\ff_q^*.$ In
other words, $f(x) - \alpha$ has exactly $r$ distinct roots.
Note that $r$ is restricted to be $r\le k-1$ since the polynomials can not
have more than $k-1$ roots.

Let $f(x)$ be a polynomial of degree $k-1$ such that it has exactly $r$
nonzero distinct roots $\alpha_1,\dots,\alpha_r$
in $\ff_q^*$.
Then $f(x)$ can be written as
$$
f(x) = \beta(x-\alpha_1) \times \cdots \times (x-\alpha_r) \times g(x),
$$
where $\beta\in\ff_q^*$
and $g(x)$ is a product of
monic irreducible polynomials, each of degree at least $2.$ The total
degree of $g(x)$ is $k-1-r.$
Since $r\ge n/2$, the polynomial $f(x)$ can not attain the value
$\alpha$, where $\alpha\in\ff_q^*$, at more than $r$ different
points $x\in\ff_q^*$ since there are $q-1-r\le n/2$ points at which the
function is nonzero.
Hence, the symbol weight of the codeword represented by
$f(x)$ is exactly $r.$
We seek the asymptotic exponent of the number of such polynomials $f(x)$.
This number is dominated by the number of possible monic irreducible
polynomials $g(x)$. To describe this
number, we recall the definition of the M\"obius function $\mu(t)$,
$$
\mu(t) \triangleq \begin{cases}
    1, & \text{if } t = 1,\\
    (-1)^s, & \text{if } t\text{ has } s\text{ distinct prime factors,}\\
    0, & \text{if } p^2|t \text{ for some prime } p.
    \end{cases}
$$
The number of monic irreducible polynomials of degree $t$ is given by the
sum
$
\frac1t\sum_{s|t, s\ge 1} \mu(s) q^{t/s}
$ (see Lidl and Niederreiter \cite[Theorem~3.25]{lid97}).
In particular, for large $t$, this sum is dominated by just the first
term $\frac1t\mu(1)q^t = \frac1t q^t.$ As a consequence, asymptotically the number of polynomials
$f(x)$ is described by the number of monic irreducible polynomials $g(x)$ of
degree $k-1-r.$ The asymptotic exponent of this count is approximately
$\lim_{n\to\infty}(k-1-r)/n = 1 - \delta- \rho$. Thus the subset $\cc'$ of
$\cc$ consists of all the codewords obtained from at least these
polynomials. Hence the count above provides a lower bound on the rate of $\cc'$:
$$
\liminf_{n\to\infty} \frac1n \log_q |\cc'| \ge 1 - \delta-\rho.
$$
The upper bound on the rate of $\cc'$ is obtained by applying both Lemma
\ref{lem:wt-distr} and Proposition~\ref{prop:sw-le-wt-distr}. We get
\begin{align*}
    |\cc'| \le |S(r)| &\le q(q-1) B_{n-r} \\
           &< q(q-1)\binom n{n-r} q^{k-r},
\end{align*}
and,
$$
\limsup_{n\to\infty} \frac 1n \log_q |S(r)| \le 
    \lim_{n\to\infty} \frac{k-r}n = 1-\delta-\rho.\quad\qquad\qed
$$

%\noindent\textsc{Proof of Proposition~\ref{prop:sw-le-wt-distr}:}
\begin{IEEEproof}[Proof of Proposition~\ref{prop:sw-le-wt-distr}]
Let $\bfc = (c_1,\dots,c_n)$  be a codeword in the Reed-Solomon code. Then
$\bfc$ is the image of a polynomial $c(x)$ evaluated at all points of
$\ff_q^* = \ff_q \setminus \{0\}$. We write $\bfc=(c(x))_{x\in\ff_q^*}.$
If $\bfc$ resulting from $c(x)$ has symbol weight exactly
$r$ then so do the codewords obtained from the polynomials $\gamma c(x) + \beta,$ for $\gamma \in
\ff_q^*, \beta\in\ff_q.$ Consider the subset $S'(r)$ of $S(r)$
that is obtained by retaining exactly one monic polynomial from the set
$\{\gamma c(x) + \beta: \gamma\in \ff_q^*, \beta\in\ff_q\}$ for any 
polynomial $c(x)$. Thus, the size of $S'(r)$ satisfies $|S'(r)|
= |S(r)|/(q(q-1))$.

We claim that $|S'(r)| \le B_{n-r}$. To show this, we claim that there
exists an
injection mapping from $S'(r)$ to the set of all
vectors of weight $n-r$. Since any $c(x)$ in
$S'(r)$ has symbol weight exactly $r$, there exists a $\beta\in\ff_q$ such
that $c(x)-\beta$ has exactly $r$  distinct roots. Thus, the codeword 
$(c(x)-\beta)_{x\in\ff_q^*}$ has Hamming weight exactly $n-r.$ This is
the only such vector. 
If there exists $e(x)\in S'(r)$ and $\alpha\in\ff_q$ such that
$(e(x)-\alpha)_{x\in\ff_q^*} = (c(x)-\beta)_{x\in\ff_q^*}$, 
then the two polynomials $c(x)$ and $e(x)$
must satisfy the relation $c(x)-\beta = e(x)-\alpha$ since they are the same
on $n=q-1$ points and their degrees are at most $k-1<n$. Thus, $c(x)
= e(x)-\alpha + \beta$, which is not possible since $S'(r)$ contains
exactly one polynomial of this form.%\hfill \qed
\end{IEEEproof}
\vspace{1mm}
%\noindent\textsc{Proof of Lemma~\ref{lem:wt-distr}:}
\begin{IEEEproof}[Proof of Lemma~\ref{lem:wt-distr}]
The expression for the weight distribution satisfies
(see \cite[Chapter  11]{mac91}):$$
B_{n-r} = \binom n{n-r} \sum_{j=0}^{k-r-1} (-1)^j \binom{n-r}j (q^{k-r-j}-1).
$$
Retaining only the first term gives the required upper bound on
$B_{n-r}$. The above expression can be rewritten as
\begin{multline*}
B_{n-r} = \binom n{n-r}\Bigg\{(q^{k-r}-1) - \sum_{i=1}^{\floor{\nicefrac{(k-r-1)}2}}
            \Bigg[ \binom{n-r}{2i-1}\times\\
            (q^{k-r-(2i-1)}-1)
        - \binom{n-r}{2i}(q^{k-r-2i}-1) \Bigg] - \\
        {I}(2\nmid k-r-1)\binom{n-r}{k-r-1}(q-1)\Bigg\},
\end{multline*}
where $I(2\nmid k-r-1)$ is an indicator function that is
1 if $2$ does not divide $k-r-1$ and 0 otherwise. We show that each of the terms in the
summation above is positive and hence we can upper bound $B_{n-r}$ by only
the first term. This is proved via the following sequence of inequalities.
For any $j,\, j=0,\dots,k-r-1$, we have
\begin{IEEEeqnarray*}{crl}
& \binom{n-r}j (q^{k-r-j}-1)\ &> \binom{n-r}{j+1} (q^{k-r-j-1}-1)\\
\Leftrightarrow& q^{k-r-j} - 1\ &> \frac{n-r-j}{j+1} (q^{k-r-j-1}-1) \\
\Leftrightarrow & q\ &> \frac{n-r-j}{j+1}\
                        \frac{1-q^{-(k-r-j-1)}}{1-q^{-(k-r-j)}}.
\end{IEEEeqnarray*}
We use the inequality $q
> (n-r)/(1-q^{-1})$. This expression is greater than the RHS of the above
because of the inequalities $1-q^{-1}
\le 1 - q^{-(k-r-j)}$, $1\ge 1-q^{-(k-r-j-1)}$, and $n-r \ge (n-r-j)/(j+1)$.
This proves the lemma.
\end{IEEEproof}
%\hfill\qed

\subsection{Discussion} % (fold)
\label{sub:Discussion}
For correcting narrowband noise in the powerline
channel, it is desirable that the symbol weight be close to the
minimum possible value of $\ceil{n/q}$. It remains open to determine
a large subset of the Reed-Solomon code for the case $\rho<1/2$.
It follows from Proposition \ref{prop:sw-le-wt-distr}
that the rate of this subset can not exceed the GV-type bound
$1-\rho-\delta$. The more interesting question is whether this bound can be
achieved, especially in the cases where $r$ is small. The work of Konyagin and Pappalardi \cite{kon06} gives an
affirmative answer to this question in the case of $r=1,$ and for
low relative distance $\delta.$ They show that
the number of permutation polynomials of degree at most $q-1-d$ is
approximately $q!/q^d$ for $d\le 0.03983q$. This is asymptotically, $\frac 1q \log_q (q!/q^d)
\simeq 1-\delta,$ where $d = \delta n$.

For the other ranges of $r$, when $r$ is growing with $n,$ we believe that
it should be possible to attain the GV bound. For instance, it would be
interesting to prove that for large $r, r\ge(k-1)/2$, and for any
irreducible polynomial $g(x)$ of degree $k-r-1$, there exists $r$ distinct
and nonzero points $\alpha_1,\dots, \alpha_r$ in $\ff_q^*$ such that the
polynomial $(x-\alpha_1)\cdots(x-\alpha_r)g(x)$ has symbol weight exactly
$r.$ Since most of the count of polynomials attaining constant symbol
weight $r$ comes from the count of irreducible polynomials, proving this
will show that the rate attains the GV-type bound.
We have obtained no counterexample on performing an exhaustive computer search over all such irreducible
polynomials $g(x)$ of degree $k-1-r$, with $(k-1)/2\le r < k<n$ in all
finite fields up to $\ff_{17}$. We are unable
to verify for larger fields because the computations become prohibitive.
We believe that the conjecture is not true for very small $r$ (when $r$
does not grow with $n$). For instance for $r=1$ the
polynomial $(x-\alpha)g(x)$, where $g(x)=x^3 + 2$ is an irreducible polynomial in
$\ff_{7}$, has symbol weight greater than 1 for every choice of
$\alpha\in\ff_{7}^*$.

% section constructions (end)

\section{Conclusion} % (fold)
\label{sec:Conclusion}

We derive the asymptotic estimates of the sizes of
 symbol weight codes. We also provide means of obtaining lower
bounds on such codes and show that it is possible to provide symbol weight
codes with the minimal possible symbol weight via recursive constructions,
given we start with a known such code. Finally, we provided
constructions of asymptotically good constant symbol weight codes.
It remains open to determine families of codes with positive rate and positive
relative distance with symbol weights that are optimal or close to
the optimal value of $\ceil{n/q}$.

% section Conclusion (end)

\section*{Acknowledgement} % (fold)
\label{sec:Acknowledgement}
We thank the Associate Editor Navin Kashyap for pointing out an error in an
earlier version of the manuscript, and for his comments, which helped
us improve the presentation of this article.
% section Acknowledgement (end)

\remove{
\begin{IEEEbiographynophoto}{Yeow Meng Chee}
(SM~'08) received the B.Math. degree in computer science and combinatorics and
optimization and the M.Math. and Ph.D. degrees in computer science, from the University of Waterloo, Waterloo, ON, Canada, in 1988, 1989,
and 1996, respectively.

Currently, he is an Associate Professor at the Division of Mathematical
Sciences, School of Physical and Mathematical Sciences, Nanyang
Technological University, Singapore. Prior to this, he was Program Director
of Interactive Digital Media R\&D in the Media Development Authority of
Singapore, Postdoctoral Fellow at the University of Waterloo and IBM's
Z{\"u}rich Research Laboratory, General Manager of the Singapore Computer
Emergency Response Team, and Deputy Director of Strategic Programs at the
Infocomm Development Authority, Singapore. His research interest lies in
the interplay between combinatorics and computer science/engineering,
particularly combinatorial design theory, coding theory, extremal set
systems, and electronic design automation.
\end{IEEEbiographynophoto}
\begin{IEEEbiographynophoto}{Han Mao Kiah}
received the B.Sc.(Hon) degree in mathematics from the National
University of Singapore, Singapore in 2006.  Currently, he is working
towards his Ph.D. degree at the Division of Mathematical Sciences, School of
Physical and Mathematical Sciences,  Nanyang Technological University, Singapore.

His research interest lies in the application of combinatorics to
engineering problems in information theory. In particular, his interests
include combinatorial design theory, coding theory and power line
communications.
\end{IEEEbiographynophoto}
\begin{IEEEbiographynophoto}{Punarbasu Purkayastha}
received the B.Tech. degree in electrical engineering from Indian
Institute of Technology, Kanpur, India in 2004, and the Ph.D. degree in
electrical engineering from University of Maryland, College Park, U.S.A.,
in 2010.

Currently, he is a Research Fellow at the Division of Mathematical
Sciences, School of Physical and Mathematical Sciences,  Nanyang
Technological University, Singapore. His research interests include coding
theory, combinatorics, information theory and communication theory.
\end{IEEEbiographynophoto}
\vfill
}
\end{document}